\documentclass{article}
\usepackage{spconf,amsmath,graphicx}
\usepackage{subcaption}
\usepackage{cite}
\usepackage{setspace}
\usepackage{amsmath}
\usepackage{amssymb}
\usepackage{amsthm}
\usepackage{mathtools}

\usepackage{array}
\usepackage{booktabs}
\usepackage{color}
\usepackage{algorithmicx}
\usepackage{algorithm}
\usepackage{algpseudocode}
\usepackage{multirow}

\title{Subsampling of Correlated Graph Signals}
\name{Rishabh Ravi$^*$, Kaushani Majumder$^*$, Kalp Vyas, and Satish Mulleti\thanks{Equal contribution}}
\address{\textit{Department of Electrical Engineering, Indian Institute of Technology (IIT) Bombay, Mumbai, India} \\
rishabhr0926@gmail.com, kaushanihp18@gmail.com, kalp.s.vyas@gmail.com, mulleti.satish@gmail.com}
\begin{document}
\ninept
\maketitle
\begin{abstract}

Graph signals are functions of the underlying graph. When the edge-weight between a pair of nodes is high, the corresponding signals generally have a higher correlation. As a result, the signals can be represented in terms of a graph-based generative model. The question then arises whether measurements can be obtained on a few nodes and whether the correlation structure between the signals can be used to reconstruct the graph signal on the remaining nodes. We show that node subsampling is always possible for graph signals obtained through a generative model. Further, a method to determine the number of nodes to select is proposed based on the tolerable error. A correlation-based greedy algorithm is developed for selecting the nodes. Finally, we verify the proposed method on different deterministic and random graphs, and show that near-perfect reconstruction is possible with node subsampling.

\end{abstract}
\begin{keywords}
Graph signal sampling, generative model, sub-Nyquist sampling, graph signal processing, greedy algorithm
\end{keywords}
\section{Introduction}
\label{sec:intro}
Sampling plays a key role in the digital processing of analog signals, and the process is realized in practice via an analog-to-digital converter (ADC). Typically, the power consumption and the cost of an ADC are proportional to the sampling rate. Hence, reducing the rate below the conventional Nyquist rate limit is desirable. This is typically achieved by exploiting the structure of the analog signal beyond bandlimitedness, such as signals in shift-invariant spaces, sparse signals, finite-rate-of-innovation signals, and more \cite{eldar_2015sampling}. Most of these results are developed for single-channel signals. In applications where several signals are measured simultaneously through multiple channels, additional structures may appear, which may be useful in reducing the overall sampling rate. 

For example, in \cite{mulleti_multiSubnyq_2021}, the authors have considered a multichannel blind-deconvolution problem and shown that the sampling rate can be reduced by using the correlation among the signals from different channels. Ahmed and Romberg  \cite{ahmed_romberg_2015, Ahmed2020Compressive} considered the problem of lowrate sampling of correlated signals. The correlation among the signals stems from each signal being represented as a linear combination of a few basis functions. As a result, the sampling matrix is shown to have a low rank, allowing the use of matrix recovery algorithms. Precisely, the matrix can be recovered from a few of its random subsamples, provided that its energy is well spread in both dimensions. Frameworks to spread energy using filters and random demodulators have also been proposed \cite{ahmed_romberg_2015, Ahmed2020Compressive}.

The aforementioned works use a specific correlation structure among the signals to lower the sampling rate. In addition, the approaches in \cite{ahmed_romberg_2015, Ahmed2020Compressive} rely on random subsampling across both the time and channel domains, which may not be practically feasible. Alternative approaches to subsampling of multiple-correlated signals were suggested in the context of sensor selection \cite{Joshi2009Sensor, MacKay1992Information, Hashemi2021Randomized, Mulleti2020Fast, Majumder2023Clustered}
and graph subsampling. In the former case, multiple sensor measurements are used to estimate some low-dimensional features of the signals. The higher the number of measurements, the better the estimation accuracy. However, this increases the computational cost as well as the power consumption. The objective of sensor selection is to select fewer sensors from a large set without reducing the estimation accuracy considerably. Note that sensor selection is possible as the dimension of the unknown features is much smaller than the number of sensors, and hence, there is always a certain amount of redundancy or correlation among the sensor measurements. Again, the correlation model varies with the applications, and there is no uniform way of characterizing it.

As with the sensor selection problem, the concept of subsampling has been extensively studied in graph signal processing \cite{Leus2024Finding, anis2014towards,sakiyama2019eigendecomposition,tanaka2020sampling,yang2021efficient, Sawarkar2024Problems}. Graphs are complex structures consisting of a set of nodes and a set of weights between each pair of nodes. Each node has an associated signal, and the correlation between any two signals is a function of the weights between those two nodes. In other words, the graph structure carries information about the correlation of the signals. Using this principle, several graph-learning strategies have been discussed where a graph (essentially the node weights) is learned from the node signals \cite{Xia2021Graph, Egilmez2017Graph, Kang2020Robust}. 

On the other hand, when a graph is given, its signals are generated to capture the underlying structure and have a certain amount of redundancy or correlation. The correlation allows for the discarding of signals from some nodes, and if required, they can be perfectly reconstructed from the signals from the remaining nodes. The correlation is typically quantified as a low-dimensional representation of the graph signals in known bases. Much research has gone into the choices of bases or low-dimensional representations \cite{chen2015signal,chen2016signal,qiu2017time,ortega2018graph}. In addition, several works discuss the design of a subsampling strategy or a way of choosing the nodes to be omitted for a given basis representation \cite{anis2014towards,sakiyama2019eigendecomposition,tanaka2020sampling,yang2021efficient, Leus2024Finding, Sawarkar2024Problems}. These methods include deterministic sampling strategies based on convex optimization \cite{tanaka2020sampling}, greedy methods \cite{Leus2024Finding}, and various heuristics such as node degree, node features, etc. \cite{Sawarkar2024Problems}, as well as random subsampling strategies \cite{tanaka2020sampling}.

A major drawback of the above-mentioned graph subsampling methods is the existence of the priors. For example, several works assume that the graph signals are bandlimited in the graph Fourier bases, which either restricts the class of signals for which the subsampling methods can be applied or leads to substantial reconstruction errors when the signals are not bandlimited.

In this paper, we consider the problem of graph subsampling by considering a generative model. Specifically, we use the model where the graph signal is written as a function of graph parameters (precisely, the edge weights). In principle, the model is similar to random signal generation, where a white noise signal is passed through a filter, and the output is a random signal with desired spectral properties. In our model, the graph plays the role of the filter. The signals generated by such a model are smooth over the graph, and we show that they are correlated based on how the nodes are connected. For example, if the weight between any pair of nodes is high, then the corresponding signals have a high correlation. This generative model eliminates the assumption of a prior knowledge of the low-dimensional subspace in which the signal lies. We show that a high correlation among the signals results in a low-dimensional approximation of the generator. This, in turn, allows for the subsampling or ignoring of some of the nodes. It has been proved that at least one node can always be ignored, yet the graph signal can be perfectly reconstructed. When a specified amount of error is tolerable, more than one node can be ignored. The existing node selection strategies either do not work for this model of the graph signals or are computationally expensive. Hence, a fast greedy algorithm based on the correlation structure among the nodes is proposed for selecting a subset of nodes on which measurements of the signals are obtained. Further, a reconstruction algorithm is proposed to reconstruct the signals on the remaining nodes of the graph. Finally, we verify our results on different types of deterministic and random graphs.

The paper is organized as follows. In Section~\ref{sec:prob_form}, we introduce and mathematically formulate the problem. The subsampling strategy and the reconstruction algorithm are presented in Section~\ref{sec:method}. The simulation experiments and the results are presented in Section~\ref{sec:expt}, and finally, the paper is concluded in Section~\ref{sec:concl}.

\section{Problem Formulation} \label{sec:prob_form}

Consider an undirected and weighted graph with $N$ nodes as $\mathcal{G} = \left(\mathcal{V}, \mathcal{E} \right)$. Here, $\mathcal{V} = \{1, \cdots, N\}$ denotes the nodes and $\mathcal{E} = \{w_{i,j} = w_{j, i} \in \mathbb{R}^+: i, j \in \mathcal{V}\}$ denotes the set of edges and its corresponding weight. The adjacency matrix $\mathbf{W}$ of the graph is an $N\times N$ matrix with its $(i, j)$-th entry given as $w_{i, j}$. A set of $N$ signals $\{y_n(t), n = 1, \cdots, N\}$ corresponding to each node is called graph signals. The structure of the signals, which depends on the graph, is used for subsampling along the node dimension, as discussed next.

Let the graph signals be represented as a vector function $\mathbf{y}(t): \mathbb{R} \rightarrow \mathbb{R}^N$. 
Then, let there exist a matrix $\mathbf{B} \in \mathbb{R}^{N \times N}$ such that $\mathbf{y}(t) = \mathbf{B}\mathbf{c}(t)$, where $\mathbf{c}(t)$ is a sparse coefficient vector with sparsity $P < N$ \cite{tanaka2020sampling}. Specifically, the graph signal has a low-dimensional representation in known bases $\mathbf{B}$. For example, $\mathbf{B}$ may represent the graph Fourier transform matrix, in which case, the graph signal is bandlimited on the graph with bandwidth $P$ if only the first $P$ coefficients of $\mathbf{c}(t)$ are non-zero \cite{pesenson2008sampling,anis2014towards,chen2016signal}.
The low-dimensional representation allows reduced measurements as $\mathbf{y}_s(t) = \mathbf{A}\mathbf{y}(t)$, where $\mathbf{A} \in \mathbb{R}^{M \times N}$ is a subsampling matrix with $P \leq M<N$. The matrix $\mathbf{A}$ could be dense or consist of $M$ rows of an $N \times N$ identity matrix. In the latter case, the process is called direct subsampling. 
The signal $\mathbf{y}(t)$ is determined from the subsampled measurements provided that $\text{rank}(\mathbf{AB}) > P$, since we need at least $P$ equations to reconstruct the $P$ non-zero coefficients of $\mathbf{c}(t)$. Note that for each $\mathbf{B}$, there could be several choices of subsampling matrices $\mathbf{A}$. However, for perfect reconstruction, it is required that $\mathbf{AB}$ has rank $P$.

The aforementioned sampling approach with direct subsampling implies that one can ignore signals from a few nodes and can still reconstruct them from the remaining ones. Hence, if $f_s$ (samples/sec.) is the time-domain sampling rate to perfectly reconstruct each $y_n(t)$ from their samples, then the overall sampling rate is $N\,f_s$. By using the structure, the sampling rate is reduced to $M\,f_s$.

In the following, we consider a generative model for the graph signals and show that sub-sampling is always possible with such models. In a low-dimensional model $\mathbf{y}(t)= \mathbf{B}\mathbf{c}(t)$, the focus is on the behavior of the graph signals on the measures of graph topology, which is captured by $\mathbf{B}$. For example, in the graph-bandlimited model, it is assumed that the graph-Fourier transform is sparse. In contrast, here, we consider the fact that the graph signals are a function of the graph, or specifically, for a graph $\mathcal{G}$ and the corresponding $\mathbf{y}(t)$, there exists a generative model $\mathcal{F}$ such that $\mathbf{y}(t) = \mathcal{F}\left(\mathcal{G} \right) \mathbf{c}(t)$. In fact, most graph learning algorithms, where a graph is learned from the data, use various generative models for the task (cf. \cite{ortega2018graph,dong2019learning} for an extensive review). For example, diffusion models are one of the widely used generators in graph learning, where $\mathcal{G} = \sum_{k = 1}^K \alpha_k \mathbf{S}^k$ and $\{\alpha_k \}_{k=1}^K$ are a set of coefficients \cite{mei2016signal, segarra2017network, thanou2017learning,pasdeloup2017characterization}. The matrix $\mathbf{S} \in \mathbb{R}^{N \times N}$ is a graph operator such as the normalized graph Laplacian operator $\mathbf{L} = \mathbf{D}^{-1/2} (\mathbf{D} - \mathbf{W})\mathbf{D}^{-1/2}$, where $\mathbf{D}$ is the degree matrix. The generative model is, in principle, similar to generating a signal through filtering. In this work, we make use of the generative matrix 
\begin{align}
    \mathbf{B} = \sum_{k = 1}^K \gamma_k \mathbf{L}^k,
    \label{eq:Bmat}
\end{align}
for constructing graph signals as $\mathbf{y}(t) = \mathbf{Bc}(t)$, where the entries of $\mathbf{c}(t)$ could be either deterministic or random, and $\gamma_k$ are deterministic coefficients known apriori. The question is whether one can subsample $\mathbf{y}(t)$ that is generated by the matrix in \eqref{eq:Bmat}? In the following, we show that the answer is always true.

\begin{figure}[!t]
    \centering
    \begin{subfigure}[b]{0.18\textwidth}
    \includegraphics[width=\textwidth]{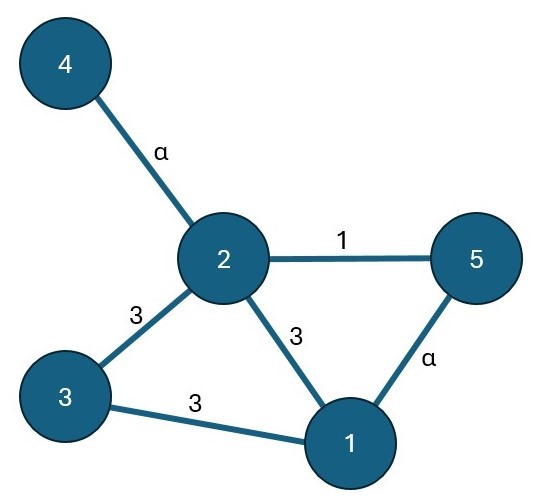}
    \caption{}
    \end{subfigure}
    \begin{subfigure}[b]{0.2\textwidth}
    \includegraphics[width=\textwidth]{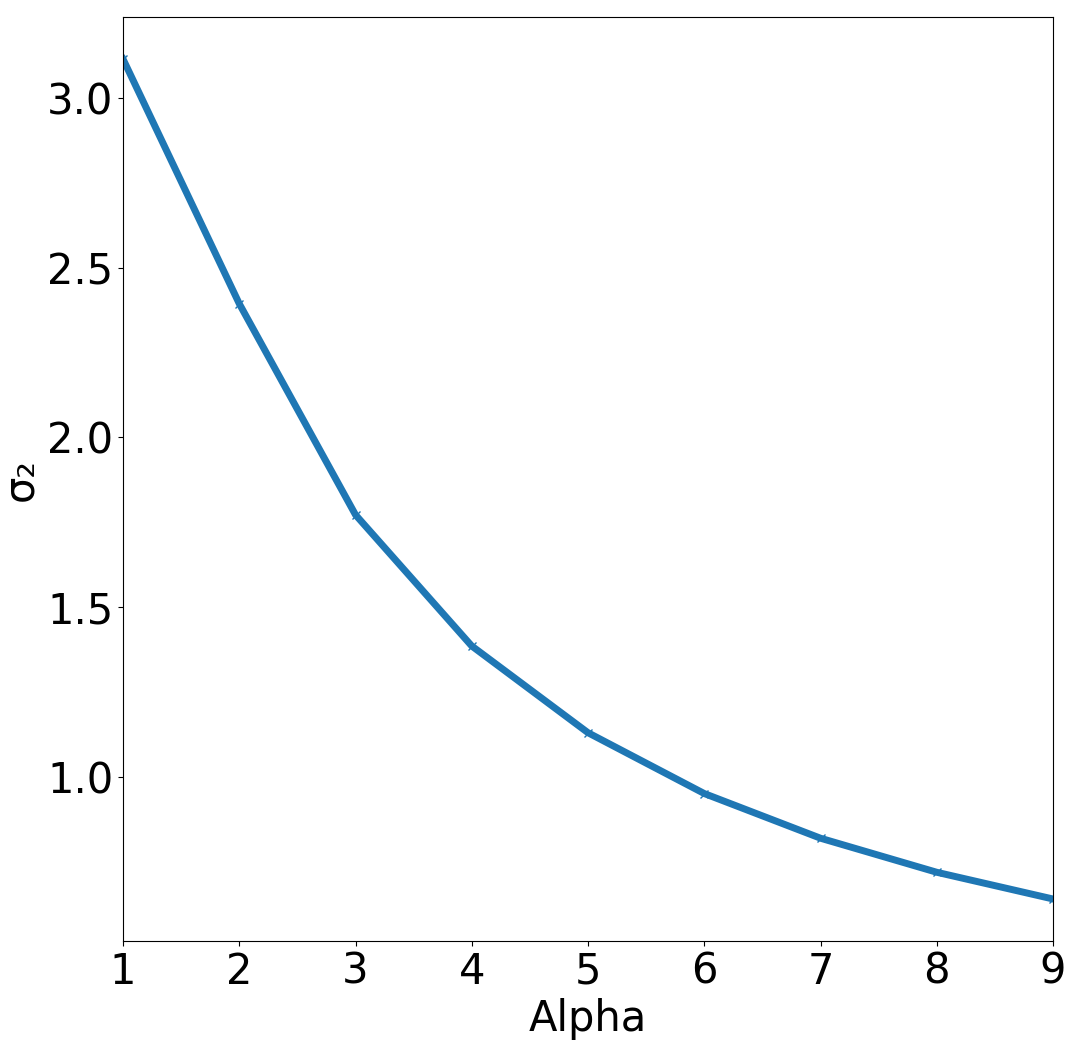}
    \caption{}
    \end{subfigure}
    \caption{Figure (a) Randomly generated weighted graph, (b) Variation of the second smallest singular value ($\sigma_2$) as a function of edge weight ($\alpha$) for the graph shown in (a)}  
    \label{fig:Grraph_correlation}
\end{figure}

\section{Graph Subsampling For Correlated Signals} \label{sec:method}

In this section, we present our proposed subsampling method, including the properties of the subsampling matrix, upper bounds on reconstruction error, and design of the subsampling matrix. We begin by discussing how the structure of a graph induces correlation among the graph signals that will allow subsampling. 

\subsection{Correlation among the graph signals}
This section discusses how the edge weights affect the correlation among the graph signals under the generative model. First, we note that the generative matrix $\mathbf{B}$ in \eqref{eq:Bmat} is rank deficient. Specifically, we have that $\text{Rank}(\mathbf{B}) \leq N-1$ as $\text{Rank}(\mathbf{L}) \leq N-1$. Let $\sigma_1\leq \sigma_2\leq \ldots \leq \sigma_N$ be the singular values, then we have that $\sigma_1 = 0$. We will show that this allows the removal of at least one node, and one can still recover the entire signal from the remaining nodes. We will discuss this shortly, but before that, let us examine if the rank can be further reduced for certain graphs. 

\begin{figure}
    \centering
    \includegraphics[width=0.7\linewidth]{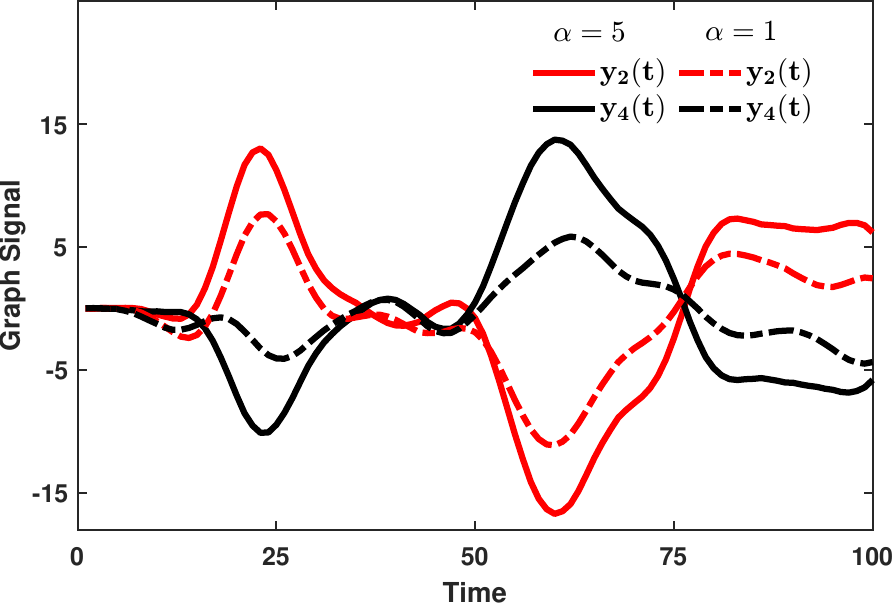}
    \caption{Plots of graph signals at Node $2$ and Node $4$ when $\alpha$ is $1$ and $5$ for graph \ref{fig:Grraph_correlation}(a)}
    \label{fig:SignalCorrelation}
\end{figure}

To this end, consider an undirected and weighted graph, as shown in Fig.~\ref{fig:Grraph_correlation}(a). Here, most nodes' weights are fixed, except for the pairs $(2, 4)$ and $(1, 5)$. For these pairs of nodes, the weights are represented by a variable $\alpha$. In Fig.~\ref{fig:Grraph_correlation}(b), the value of the second smallest singular value, $\sigma_2$, is shown as a function of $\alpha$. The larger the $\alpha$, the smaller $\sigma_2$ is, which implies that an $N-2$ rank matrix can well approximate the matrix. The low-rank tendency of the matrix $\mathbf{B}$ manifests as a correlation among the node signals. To visualize the correlation, In Fig.~\ref{fig:SignalCorrelation}, the signals corresponding to nodes two and four are shown as a function of time for $\alpha = 1$ and $\alpha = 5$. In these plots, the signals are generated as $\mathbf{y}(t) = \mathbf{Bc}(t),$ where $\mathbf{B}$ is given by \eqref{eq:Bmat}, $\mathbf{c}(t)$ are samples of some continuous functions, and $K = 5$, $\alpha_i$'s taken from a normal distribution. As expected, the corresponding signals are more correlated when the weights between any two nodes increase. Specifically, correlations $0.58$ and $0.97$, for $\alpha = 1$ and $5$, respectively. The low-rank approximation or the correlation allows subsampling, and reconstruction with some degree of error, as discussed next.

\subsection{Proposed Subsampling and Reconstruction} \label{sec:error_rec}
Our aim is to design a subsampling or node selection method for the graph signal generated using an approximate low-rank generator matrix, as described in the previous section. Before introducing our method, we will examine the reconstruction error resulting from subsampling. The bounds will serve as a basis for guiding the node selection process.

As the subsampling is invariant to time, we drop the time index from the node signals $\mathbf{y}(t)$ for brevity and denote it as a vector $\mathbf{y}$. The measurement model is given by
\begin{align}
    \mathbf{y} = \mathbf{Bc}.
    \label{eq:y_meas}
\end{align}

As discussed previously, depending upon the edge weights and connectivity, the generating matrix $\mathbf{B}$ may be approximated by a low-rank matrix of rank smaller than $N-1$. In such a scenario, the approximate measurement model may be represented as
\begin{align}
    \tilde{\mathbf{y}} = \tilde{\mathbf{B}}\mathbf{c},
    \label{eq:y_tilde}
\end{align}
where, $\tilde{\mathbf{B}} \in \mathbb{R}^{N \times N}$ is a $P$-rank approximation of $\mathbf{B}$ and correspondingly $\tilde{\mathbf{y}}$ is a projection of $\mathbf{y}$ onto a lower-dimensional subspace. Since $\text{Rank}(\tilde{\mathbf{B}}) = P$, it can be decomposed as
\begin{align}
    \tilde{\mathbf{B}} = \mathbf{T F},
    \label{eq:btilde}
\end{align}
where $\mathbf{T} \in \mathbb{R}^{N \times P}$ is a full column-rank matrix and $\mathbf{F} \in \mathbb{R}^{P \times N}$ is a coefficient matrix that generates $\tilde{\mathbf{B}}$ from $\mathbf{T}$.

If $\tilde{\mathbf{y}}$ is available at only $P$ nodes, its values at the remaining $N-P$ nodes could be reconstructed perfectly by using the low-rank nature of $\tilde{\mathbf{B}}$. Let $\mathcal{S}$ be a subset of the nodes where $\tilde{\mathbf{y}}$ is observed. We defer the discussion of choosing $\mathcal{S}$ to Section \ref{sec:subsampling_strategy}. From the discussion above, it suffices if $|\mathcal{S}|= P$. Let $\mathbf{A}_\mathcal{S} \in \mathbb{R}^{P \times N}$ denote the corresponding subsampling matrix formed by selecting the rows of a $N \times N$ identity matrix indexed by $\mathcal{S}$. 
Then, the subsampled signal $\tilde{\mathbf{y}}_\mathcal{S}$ is given by, 
\begin{align}
    \tilde{\mathbf{y}}_{\mathcal{S}} = \mathbf{A}_{\mathcal{S}} \tilde{\mathbf{y}} = \mathbf{A}_{\mathcal{S}} \tilde{\mathbf{B}} \mathbf{c} = \mathbf{A}_{\mathcal{S}} \mathbf{T} \mathbf{Fc}.
\end{align} 
If $\mathbf{A}_{\mathcal{S}} \mathbf{T} \in \mathbb{R}^{P \times P}$ is invertible, then $\tilde{\mathbf{y}}$ can be reconstructed perfectly.
\begin{align}
 {\tilde{\mathbf{y}}} = \mathbf{T} (\mathbf{A}_{\mathcal{S}} \mathbf{T} )^{-1} \tilde{\mathbf{y}}_{\mathcal{S}}.
 \label{eq:recon}
 \end{align}

In the above analysis, we have assumed that $\tilde{\mathbf{y}}_{\mathcal{S}}$ is available. However, we only have access to the subsampled graph signal $\mathbf{y}_{\mathcal{S}} = \mathbf{A}_{\mathcal{S}} \mathbf{y}$.
Using this, the reconstructed signal is
 \begin{align}
    \hat{\tilde{\mathbf{y}}} = \mathbf{T} (\mathbf{A}_S \mathbf{T} )^{-1} {\mathbf{y}}_{\mathcal{S}}.
    \label{eq:y_hat_tilde}
\end{align}
We can thus reconstruct a lower-dimensional projection of $\mathbf{y}$ from the samples obtained at only $P$ nodes. Next, we analyze the error in this reconstruction.

The reconstruction error is defined as 
\begin{align}
    \| \mathbf{y} - \hat{\tilde{\mathbf{y}}}\|_2 
   &= \| \mathbf{y} - \tilde{\mathbf{y}} +  \tilde{\mathbf{y}}-\hat{\tilde{\mathbf{y}}} \|_2 \notag \\
   &= \|(\mathbf{B} - \tilde{\mathbf{B}})\mathbf{c} +  \mathbf{T} (\mathbf{A}_S \mathbf{T} )^{-1} (\tilde{\mathbf{y}}_{\mathcal{S}} - \mathbf{y}_{\mathcal{S}})\|_2 \notag \\
   &\le \|(\mathbf{B} - \tilde{\mathbf{B}}) \mathbf{c}\|_2 + \| \mathbf{T} (\mathbf{A}_{\mathcal{S}} \mathbf{T} )^{-1} (\tilde{\mathbf{y}}_{\mathcal{S}} - \mathbf{y}_{\mathcal{S}})\|_2,
   \label{eq:err_gen}
\end{align}
where the inequality follows from the triangle inequality. 
The first term of \eqref{eq:err_gen} denotes the error due to the low-rank approximation of $\mathbf{B}$. In contrast, the second term determines the error accumulated from using $\mathbf{y}_{\mathcal{S}}$ instead of $\tilde{\mathbf{y}}_{\mathcal{S}}$ in the reconstruction. To minimize this error, we can choose an appropriate low-rank approximation of $\mathbf{B}$ and a sampling matrix $\mathbf{A}_{\mathcal{S}}$.
 
A $P$-rank approximation of $\mathbf{B}$ can be constructed by taking the singular value decomposition (SVD) of $\mathbf{B}$ and nullifying the $P$ smallest singular values. The Eckart–Young–Mirsky theorem \cite{Eckart1936Approximation} states that this $\tilde{\mathbf{B}}$ minimizes the first term of the error given by \eqref{eq:err_gen}. Such an approximation can be represented as

\begin{align}
    {\bf {\tilde B}}_{\text{svd}} = {\bf U}{\boldsymbol{\Sigma}}{\bf V}^T,
\end{align}
where $\boldsymbol{\Sigma} \in \mathbb{R}^{P \times P}$ is the diagonal matrix of its singular values and $\mathbf{U}, \mathbf{V} \in \mathbb{R}^{N \times P}$ are matrices with columns representing the left and right singular vectors of $\tilde{\mathbf{B}}_{\text{svd}}$ respectively. It is easy to see that in this case $\mathbf{T} = \mathbf{U}$ and $\mathbf{F} =\boldsymbol{\Sigma} \mathbf{V}^T$.
The proposed reconstruction method \eqref{eq:y_hat_tilde} assumes ${\bf A}_{\mathcal{S}}{\bf T}$ to be invertible. Here, since $\mathbf{U}$ is a unitary matrix with full column rank, we can find a subset $\mathcal{S}$ of its rows which are linearly independent, making ${\bf A}_{\it S}{\bf U}$ invertible. 

Using $\tilde{\mathbf{B}}_{\text{svd}}$ ensures that the first error term in \eqref{eq:err_gen} is minimized. However, the second term of the error depends on the selected set of nodes $\mathcal{S}$, and no bound can be specified for this term. This second term in \eqref{eq:err_gen} can be minimized to zero by using a $\tilde{\mathbf{B}}$ such that $\mathbf{y}_{\mathcal{S}} = \tilde{\mathbf{y}}_{\mathcal{S}}$. From \eqref{eq:y_meas} and \eqref{eq:y_tilde}, it is evident that this condition is satisfied when $\mathbf{A}_{\mathcal{S}} \mathbf{B} = \mathbf{A}_{\mathcal{S}} \tilde{\mathbf{B}}$. A $P$-rank approximation $\tilde{\mathbf{B}}$ denoted as $\tilde{\mathbf{B}}_{\text{samp}}$, that satisfies the previous equality is constructed as follows. The rows of $\tilde{\mathbf{B}}_{\text{samp}}$ indexed by the set $\mathcal{S}$ are taken to be the corresponding rows of $\mathbf{B}$. The remaining rows of $\mathbf{B}$ are projected onto the subspace spanned by the rows of $\mathbf{A}_{\mathcal{S}} \mathbf{B}$ to form the corresponding rows of $\tilde{\mathbf{B}}_{\text{samp}}$. To ensure $\tilde{\mathbf{B}}_{\text{samp}}$ is of rank $P$, the set $\mathcal{S}$ has to be selected so that the rows of $\mathbf{A}_{\mathcal{S}} \mathbf{B}$ are linearly independent.

Once $\tilde{\mathbf{B}}_{\text{samp}}$ is generated, $\mathbf{T}$ can be obtained by taking $P$ independent columns of $\tilde{\mathbf{B}}_{\text{samp}}$ and it spans the column space of $\tilde{\mathbf{B}}_{\text{samp}}$. The matrix $\mathbf{F}$ is then a coefficient matrix that generates $\tilde{\mathbf{B}}_{\text{samp}}$ from $\mathbf{T}$, and given as
\begin{align}
    {\bf F}_{:,i} =
\left\{
	\begin{array}{ll}
		{\bf e}_i,  & \mbox{if } i \in {\it S}^c, \\
		\boldsymbol{\beta}, & \mbox{if } i \in {\it S},
	\end{array}
\right.
\label{eq:F_struc}
\end{align}
where ${\bf e}_i$ are the standard basis vectors of $\mathbb{R}^{P}$ and $\boldsymbol{\beta}\in \mathbb{R}^{P}$ are some coefficient vectors such that \eqref{eq:btilde} holds. For reconstruction of the signal (cf. \eqref{eq:recon}), explicitly evaluation pf ${\bf F}$ is not required.

For the reconstruction, we require ${\bf A}_{\it S}{\bf T}$ to be invertible.
Recall that $\mathbf{A}_{\mathcal{S}} \tilde{\mathbf{B}}_{\text{samp}} \in \mathbb{R}^{P \times N}$ has full row rank and is obtained from a symmetric ${\bf B}$ by selecting $P$ rows; we can select the same $P$ columns indexed by $\mathcal{S}$ to form $\mathbf{T}$. This makes $\mathbf{A}_{\mathcal{S}} \mathbf{T}$ square and invertible.

Using $\tilde{\mathbf{B}}_{\text{samp}}$ ensures that an error bound is achieved that nullifies the error that accumulates by using $\mathbf{y}_{\mathcal{S}}$ instead of $\tilde{\mathbf{y}}_{\mathcal{S}}$, at the cost of the error in the low-rank approximation $\tilde{\mathbf{B}}$. There is not a direct result to show which low-rank approximation among $\tilde{\mathbf{B}}_{\text{svd}}$ and $\tilde{\mathbf{B}}_{\text{samp}}$ produces a smaller error. Moreover, it should be noted that if ${\bf B}$  was of rank $P$, we can easily decompose ${\bf B}$ into ${\bf TF}$ without having to create a ${\bf \tilde{B}}$, thus giving us perfect reconstruction. Also, as $\text{Rank}({\bf B})\leq N-1,$the approach always allows the removal of at least one node and has perfect reconstruction.

It should also be noted that a higher correlation implies a better lower-rank approximation is possible. The approximation level is also subjected to the choice of nodes, and an approach to the choice is discussed next.

\subsection{Node Selection for Subsampling} \label{sec:subsampling_strategy}
From section~\ref{sec:error_rec}, it is evident that the reconstruction error is dependent on the rank $P$ of the low-rank approximation $\tilde{\mathbf{B}}$ as well as the selected set of nodes $\mathcal{S}$. In this section, we discuss methods to select $P$ and $\mathcal{S}$.

First, we discuss how to select $P$ and the approach is same whether we use $\tilde{\mathbf{B}}_{\text{svd}}$ or $\tilde{\mathbf{B}}_{\text{samp}}$ as the low-rank approximation of $\mathbf{B}$. To this end, we use the SVD-based low-rank approximation of $\mathbf{B}$ and the approximation error to determine $P$. From the Eckart–Young–Mirsky theorem \cite{Eckart1936Approximation}, we note that
\begin{align}
    \left\| \mathbf{B} - \tilde{\mathbf{B}}_{\text{svd}}  \right\|_2 &= \left( \sum_{i = P+1}^N \sigma_i^2 \right)^{1/2},
    \label{eq:low_rank_error}
\end{align}
where $\sigma_i$ are the singular values of $\mathbf{B}$ arranged in a descending order. Note that \eqref{eq:low_rank_error} can always be determined for a given $P$, and we use this error metric to choose $P$.  Specifically, if the maximum tolerable normalized error for the low-rank approximation is $\varepsilon$, then $P$ can be chosen such that $\displaystyle \left( \sum_{i = P+1}^N \sigma_i^2 \right)^{1/2} \le \varepsilon \left( \sum_{i = 1}^N \sigma_i^2 \right)^{1/2}$. Here, $\epsilon$ is user-specific and affects the reconstruction error in \eqref{eq:err_gen}. 

Once $P$ is chosen, the next step is to choose the sampling set $\mathcal{S}$. For this purpose, the correlation structure of $\mathbf{B}$ or $\tilde{\mathbf{B}}$ is used. In particular, considering $\tilde{\mathbf{B}}_{\text{svd}}$ is used as the low-rank approximation of $\mathbf{B}$, $\mathcal{S}$ needs to be chosen such that $\mathbf{A}_{\mathcal{S}} \mathbf{U}$ is invertible. This will happen when $\mathcal{S}$ selects nodes that are least correlated with other nodes in $\tilde{\mathbf{B}}_{\text{svd}}$. On the other hand, the low-rank approximation $\tilde{\mathbf{B}}_{\text{samp}}$ needs the knowledge of the set $\mathcal{S}$ for its formation. In this case, the correlation among the nodes in $\mathbf{B}$ can be utilized to select $\mathcal{S}$. 

Starting from $\tilde{\mathbf{B}}$ for the SVD-based approximation and $\mathbf{B}$ for $\tilde{\mathbf{B}}_{\text{samp}}$ approximation, a greedy strategy is employed for the selection process. In every iteration, from the remaining nodes, two nodes that have the highest correlation are obtained. Among these two, one node is removed, which has an overall higher correlation with all the nodes. The steps of this selection process are presented in Algorithm~\ref{alg:NS_alg}. Once $\tilde{\mathbf{B}}$ and $\mathcal{S}$ are determined, $\mathbf{y}$ can be estimated by using the method presented in Section~\ref{sec:error_rec}.

\begin{algorithm}[tb]
    \caption{Correlation-Based Greedy Node Selection Strategy}
    \label{alg:NS_alg}
    \begin{algorithmic}[1]
        \State \textbf{Initialize:} $\varepsilon$, $\mathbf{B}_{\text{sel}} = \mathbf{B}$ or $\tilde{\mathbf{B}}$
        \State $\boldsymbol{\sigma} \coloneqq \text{SVD}(\mathbf{B})$
        \State $P = \left\{ \min\{j\} | \sum_{i=j+1}^N \sigma_i^2 \le \varepsilon^2 \| \boldsymbol{\sigma} \|_2^2 \right\}$
        \Comment{Finding $P$}

        \State $\mathbf{B}_{\text{norm}} = $ row normalized $\mathbf{B}_{\text{sel}}$
        \State $\mathbf{B}_{\text{corr}} = \text{abs} \left( \mathbf{B}_{\text{norm}}^H \mathbf{B}_{\text{norm}} \right)$
        \State $\text{diag} \left( \mathbf{B}_{\text{corr}} \right) = \mathbf{0}$

        \State $\mathcal{S} = \{1, \ldots, N\}$

        \While{$\left| \mathcal{S} \right| > P$} 
            \Comment{Select $\mathcal{S}$}
            \State $[i,j] = \arg \max \left( \mathbf{B}_{\text{corr}} [\mathcal{S}, \mathcal{S}] \right)$
            \If {sum$\left( \mathbf{B}_{\text{corr}}[i,:] \right) >$ sum$\left( \mathbf{B}_{\text{corr}}[j,:] \right)$ }
                \State $\mathcal{S} = \mathcal{S} \backslash \{i\}$
            \Else 
                \State $\mathcal{S} = \mathcal{S} \backslash \{j\}$
            \EndIf
        \EndWhile
    \end{algorithmic}
\end{algorithm}

In the next section, we validate our proposed method of graph subsampling and reconstruction on different graphs.

\section{Experiments and Results} \label{sec:expt}
In this section, we validate our method for different graphs. The graph signals are generated using a $\mathbf{B}$ matrix given by \eqref{eq:Bmat}, and $\mathbf{c}(t)$ across the different nodes generated randomly from a normal distribution and varying smoothly with $t$. The low-rank representation of $\mathbf{B}$ is calculated using both $\tilde{\mathbf{B}}_{\text{svd}}$ and $\tilde{\mathbf{B}}_{\text{samp}}$ as described in section~\ref{sec:error_rec}. Node selection is performed for each of these cases using Algorithm~\ref{alg:NS_alg}. Tables~\ref{tab:Res_Graph1_alpha_5} and \ref{tab:Res_Graph1_alpha_1} present the results for the graph shown in fig.~\ref{fig:Grraph_correlation}(a) for $\alpha = 5$ and $1$ respectively and for $\varepsilon = 0.03$ and $0.01$. As expected, when $\varepsilon$ decreases, the number $P$ of nodes to select increases, and the error in reconstruction decreases. In particular, the error is $-18$ to $-19$ dB when $\varepsilon = 0.03$, and it decreases by around $10$ dB when $\varepsilon = 0.01$. For a given $\varepsilon$, using $\tilde{\mathbf{B}}_{\text{samp}}$ gives marginally better reconstruction than $\tilde{\mathbf{B}}_{svd}$ in terms of the normalized error. Moreover, for the case when $\alpha = 1$ and $\varepsilon = 0.01$, $P = 4$ nodes need to be selected as is evident from Table~\ref{tab:Res_Graph1_alpha_1} and in this case perfect reconstruction occurs. This is expected since $\mathbf{B}$ is of rank $4$, and hence, the fifth signal can be reconstructed perfectly from the signals of the selected $4$ nodes.

We further validate the proposed method for different types of graphs, such as an Erdos-R\`enyi graph with $p = 0.5$, a complete graph, and a bipartite graph. In each case, the edge weights are selected from a uniform distribution in $[1,10]$. The results are presented in Tables~\ref{tab:Res_Erdos_Renyi}, \ref{tab:Res_Comp_Graph}, and \ref{tab:Res_Bipartite} respectively. The results show similar trends as in Tables~\ref{tab:Res_Graph1_alpha_5} and \ref{tab:Res_Graph1_alpha_1}. Further, it is observed that selecting fewer number of nodes produces similar errors when there is more correlation among the nodes of the graph.

\section{Conlusion} \label{sec:concl}

In this paper, a node subset selection algorithm is proposed for graph signals generated using a linear combination of different moments of the graph Laplacian. Theoretically, it was shown that for such graph signals, at least one node can always be removed, and still, perfect reconstruction is possible from the subsampled signal. When some error is tolerable, further reduction is possible in the number of nodes selected. The error of the reconstructed signal is derived, and a subsampling strategy is proposed that depends on the correlation between the nodes. We verify our results on different types of graphs.

\begin{table*}
    \centering
    \caption{Results for Graph $1$(a) with $\alpha = 5$.}
    \label{tab:Res_Graph1_alpha_5}
    \bgroup
    \renewcommand{\arraystretch}{1.5}
    \begin{tabular}{|m{0.2in}|m{0.25in}|m{1.6in}|m{3.3in}|m{0.75in}|}
        \hline
        $\boldsymbol{\varepsilon}$ & $\tilde{\mathbf{B}}$ & \textbf{Selected Nodes (in green)} & \textbf{Original and Reconstructed Signals} & \textbf{Normalized Error (in dB)} \\
        \hline
        \hline
        \multirow{4}{*}{0.03} &&&& \\
        &$\tilde{\mathbf{B}}_{\text{svd}}$ &  \includegraphics[width=1.6in]{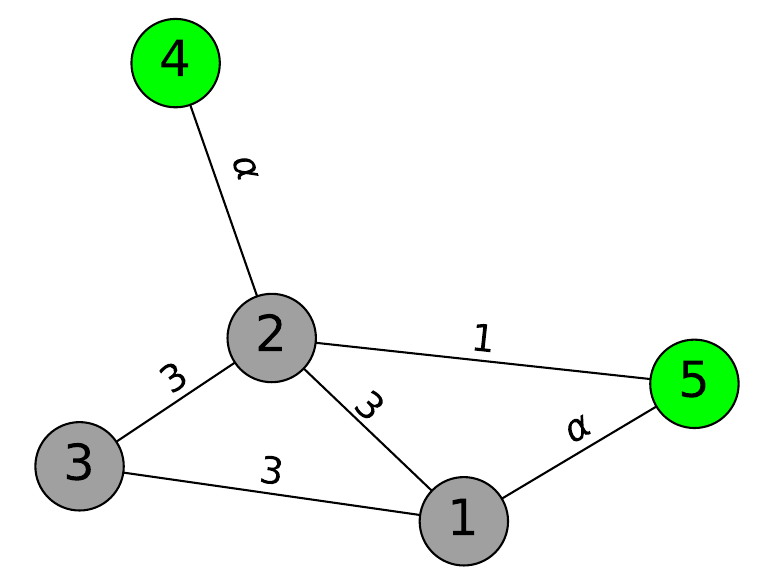} & \includegraphics[width=3.3in]{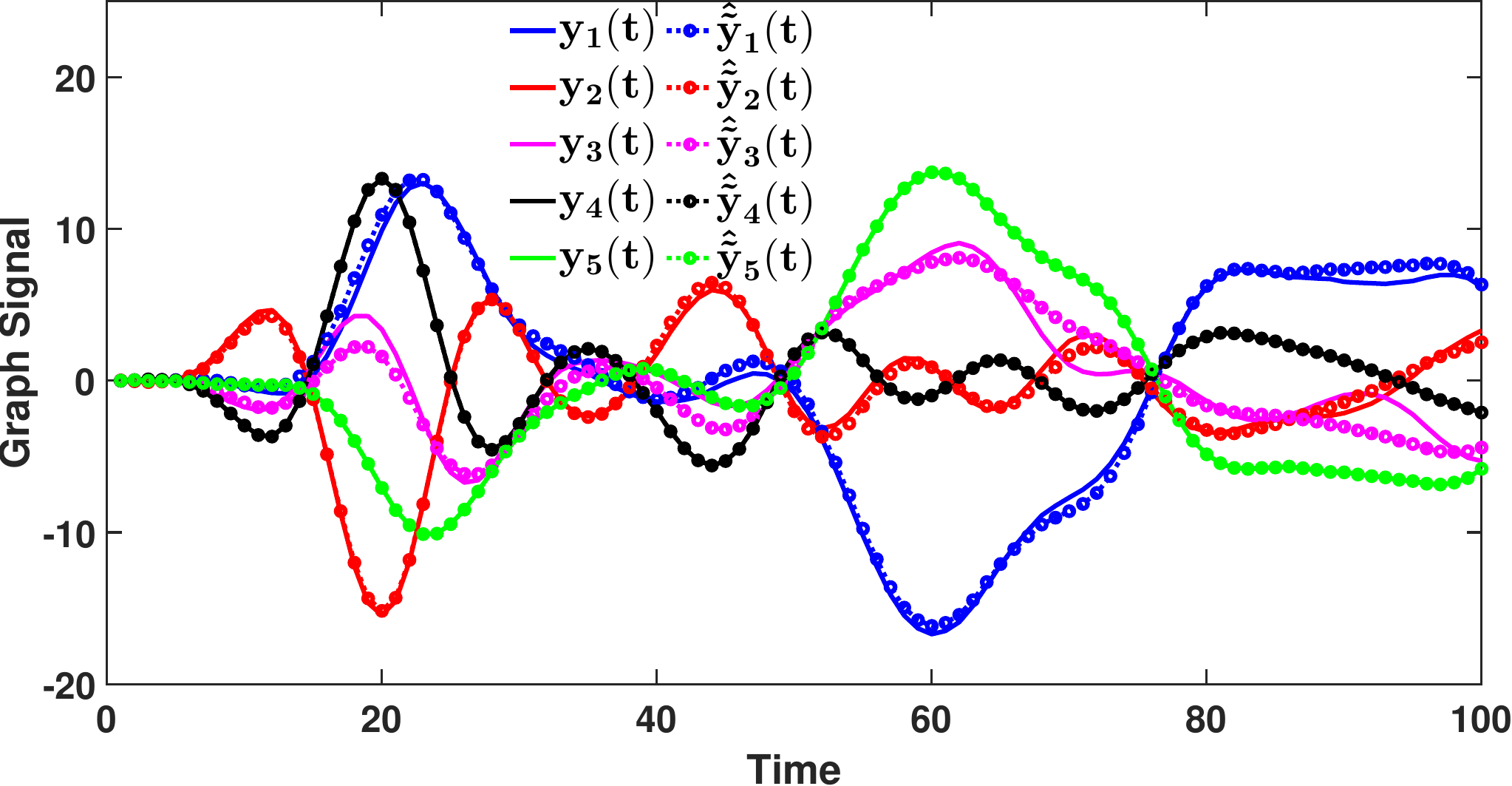} & $-18.83$ \rule{0pt}{4ex} \\
        \cline{2-5}
        &&&& \\
        & $\tilde{\mathbf{B}}_{\text{samp}}$ &  \includegraphics[width=1.6in]{ICASSP_format/Images/Graph1_P_2_alpha_5.pdf} & \includegraphics[width=3.3in]{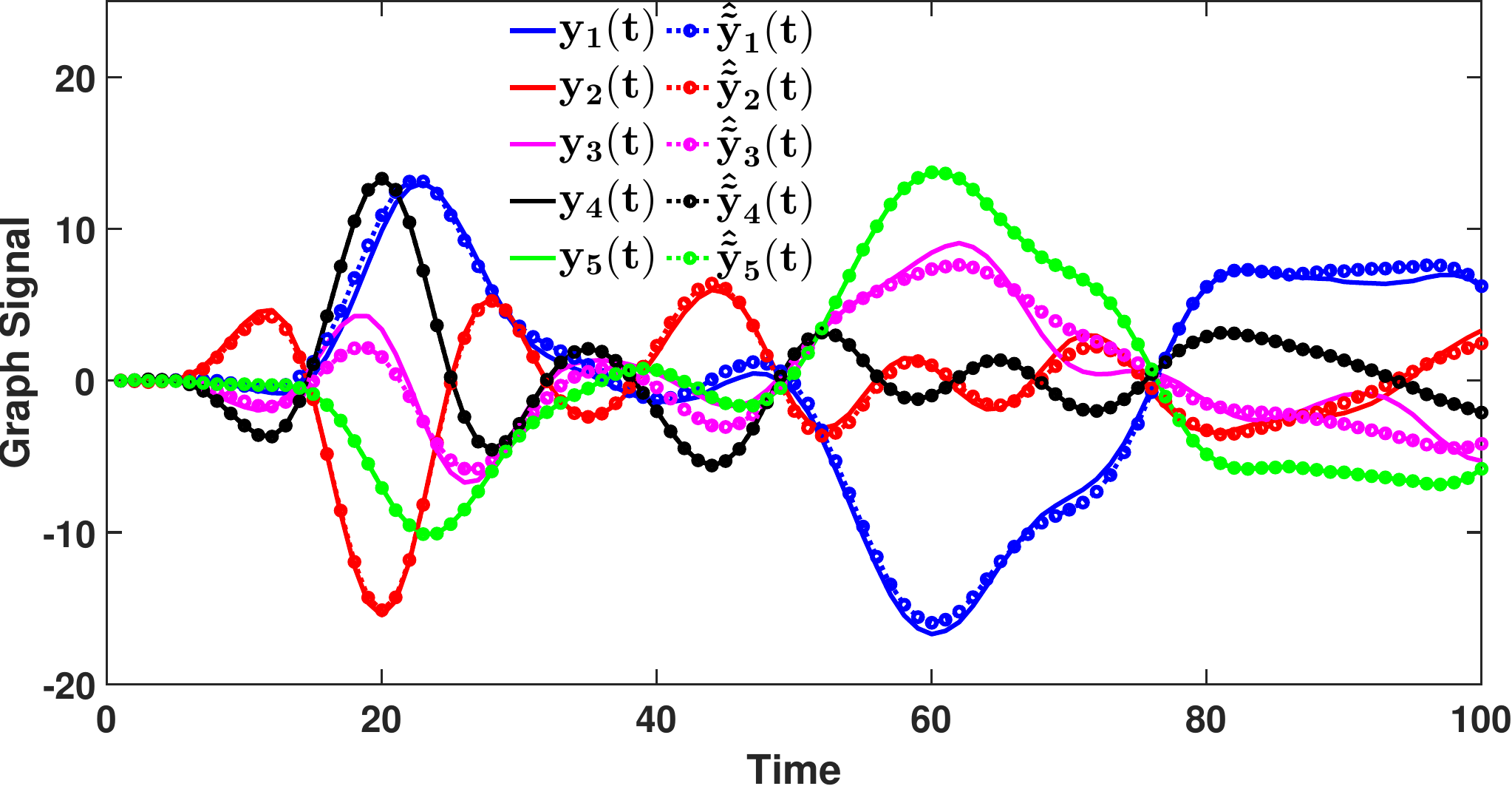} & $-19.09$ \\
        \hline
        \multirow{2}{*}{0.01} &&&& \\
        &$\tilde{\mathbf{B}}_{\text{svd}}$ &  \includegraphics[width=1.6in]{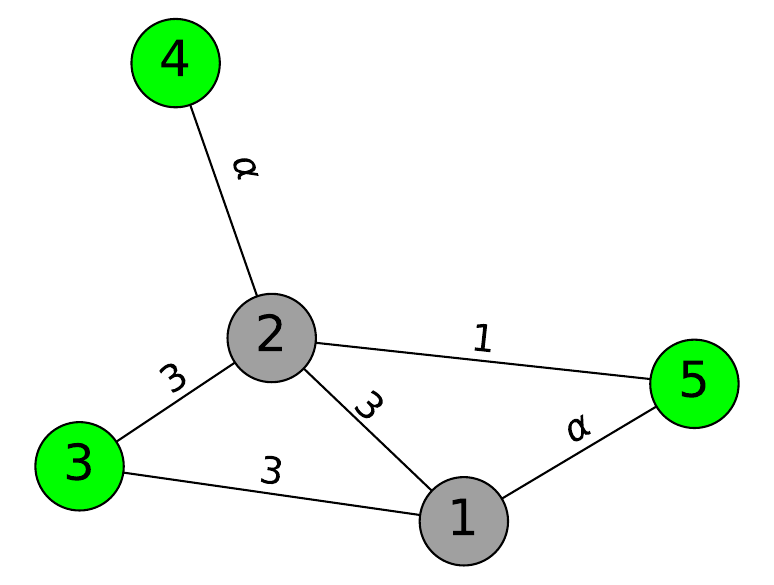} & \includegraphics[width=3.3in]{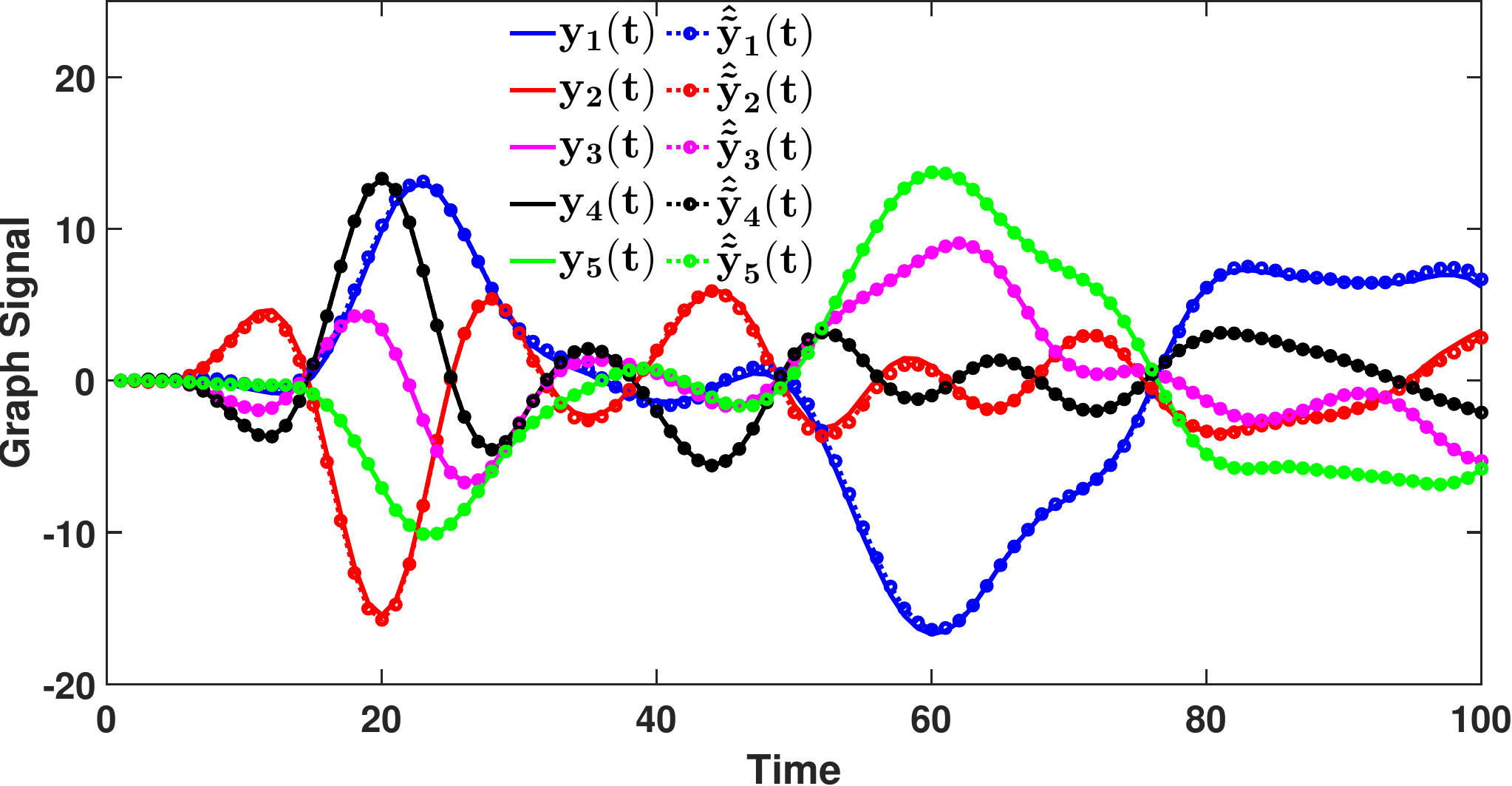} & $-29.26$ \\
        \cline{2-5}
        &&&& \\
        & $\tilde{\mathbf{B}}_{\text{samp}}$ &  \includegraphics[width=1.6in]{ICASSP_format/Images/Graph1_P_3_alpha_5.pdf} & \includegraphics[width=3.3in]{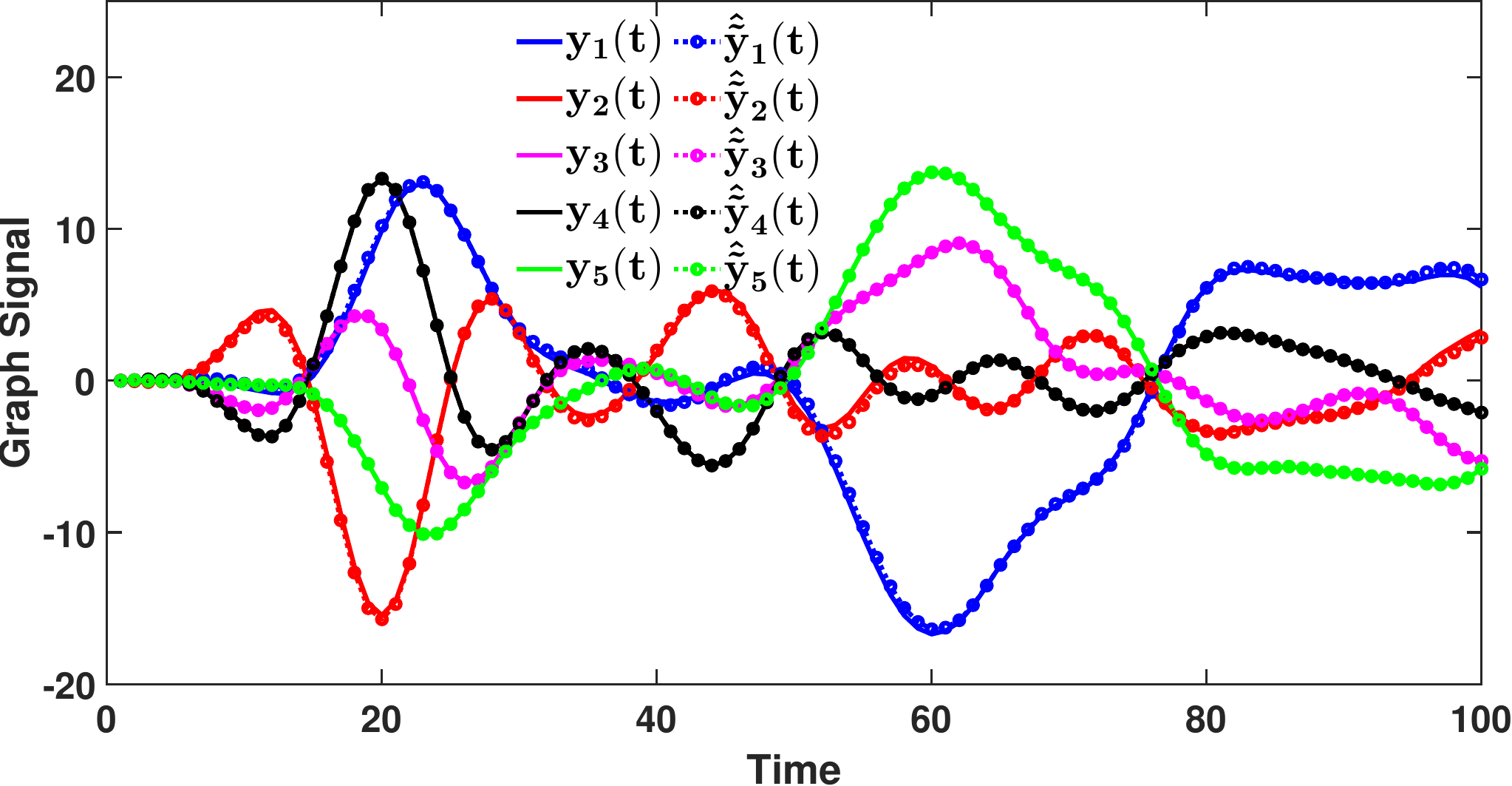} & $-29.28$ \\
        \hline

    \end{tabular}
    \egroup
\end{table*}

\begin{table*}
    \centering
    \caption{Results for Graph $1$(a) with $\alpha = 1$.}
    \label{tab:Res_Graph1_alpha_1}
    \bgroup
    \renewcommand{\arraystretch}{1.5}
    \begin{tabular}{|m{0.2in}|m{0.25in}|m{1.6in}|m{3.3in}|m{0.75in}|}
        \hline
        $\boldsymbol{\varepsilon}$ & $\tilde{\mathbf{B}}$ & \textbf{Selected Nodes} & \textbf{Original and Reconstructed Signals} & \textbf{Normalized Error (in dB)} \\
        \hline
        \hline
        \multirow{4}{*}{0.03} &&&& \\
        &$\tilde{\mathbf{B}}_{\text{svd}}$ &  \includegraphics[width=1.6in]{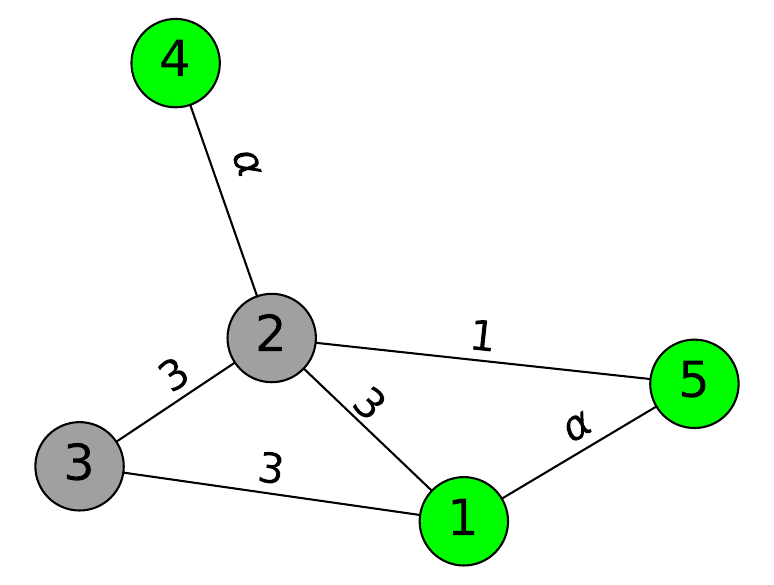} & \includegraphics[width=3.3in]{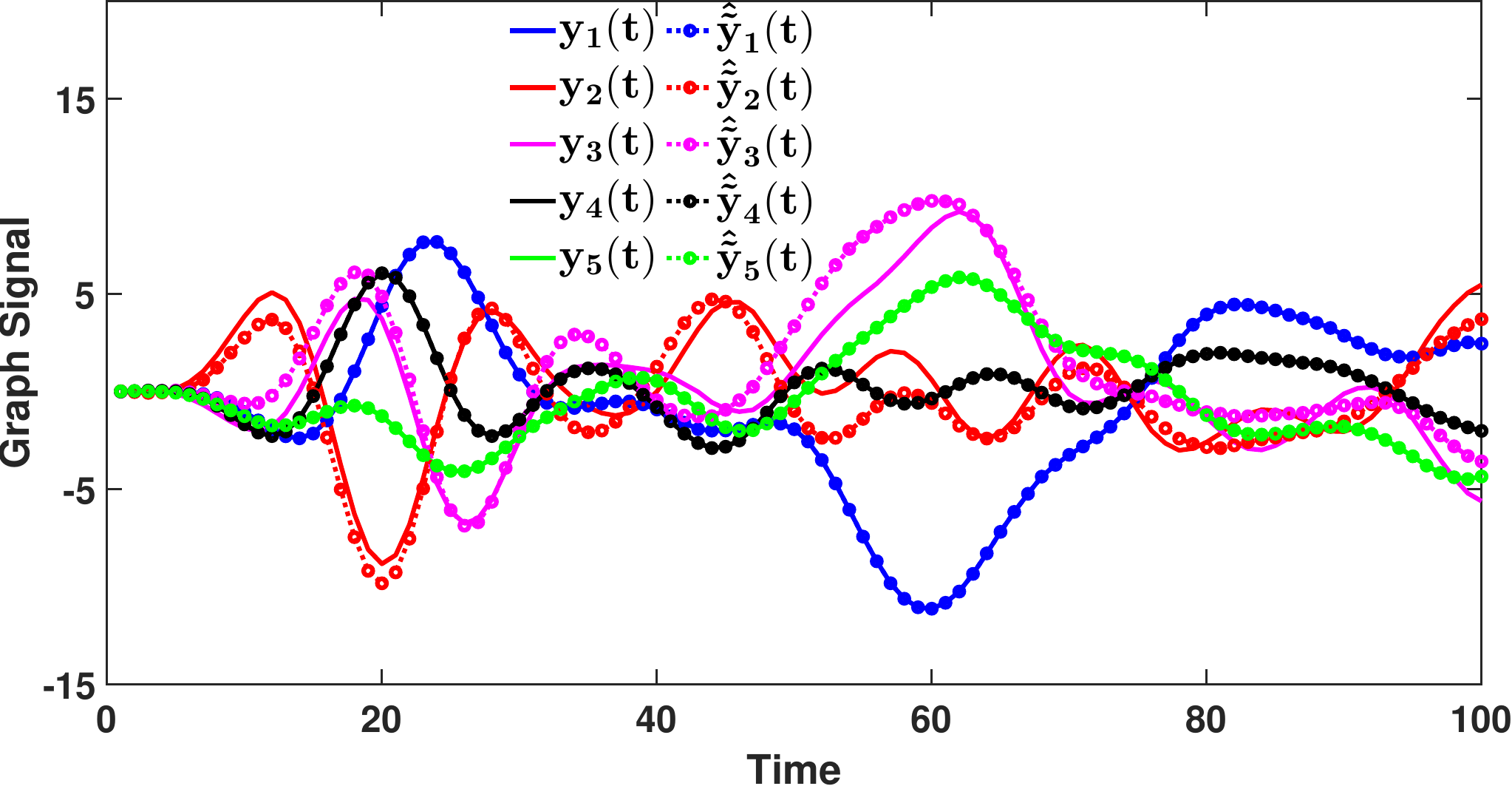} & $-12.16$ \\
        \cline{2-5}
        &&&& \\
        & $\tilde{\mathbf{B}}_{\text{samp}}$ &  \includegraphics[width=1.6in]{ICASSP_format/Images/Graph1_P_3_alpha_1.pdf} & \includegraphics[width=3.3in]{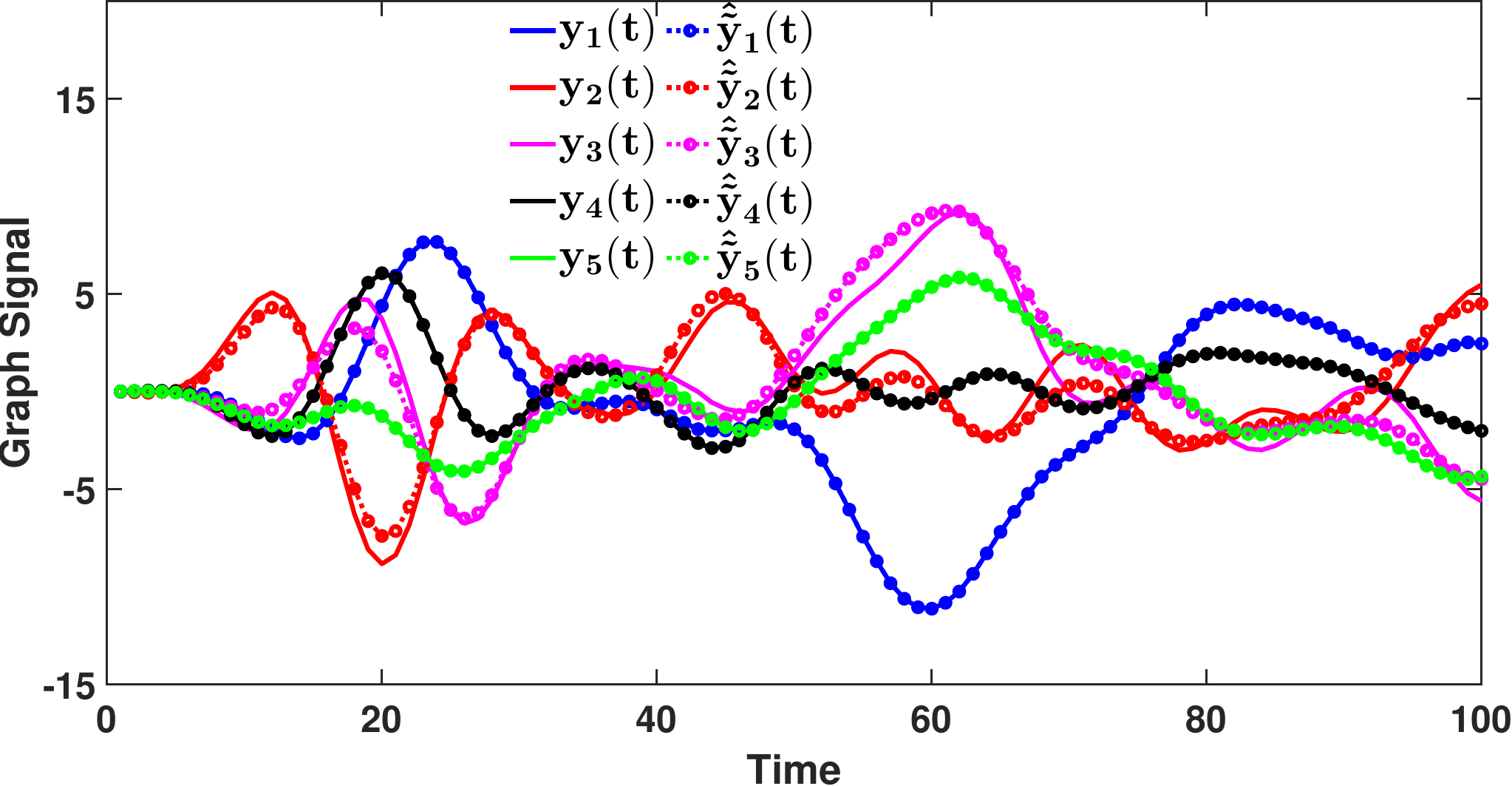} & $-14.95$ \\
        \hline
        \multirow{2}{*}{0.01} &&&& \\
        &$\tilde{\mathbf{B}}_{\text{svd}}$ &  \includegraphics[width=1.6in]{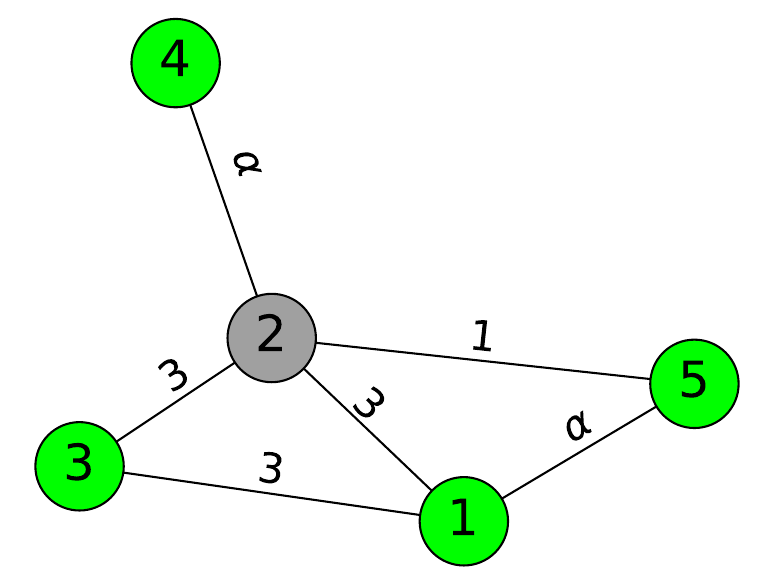} & \includegraphics[width=3.3in]{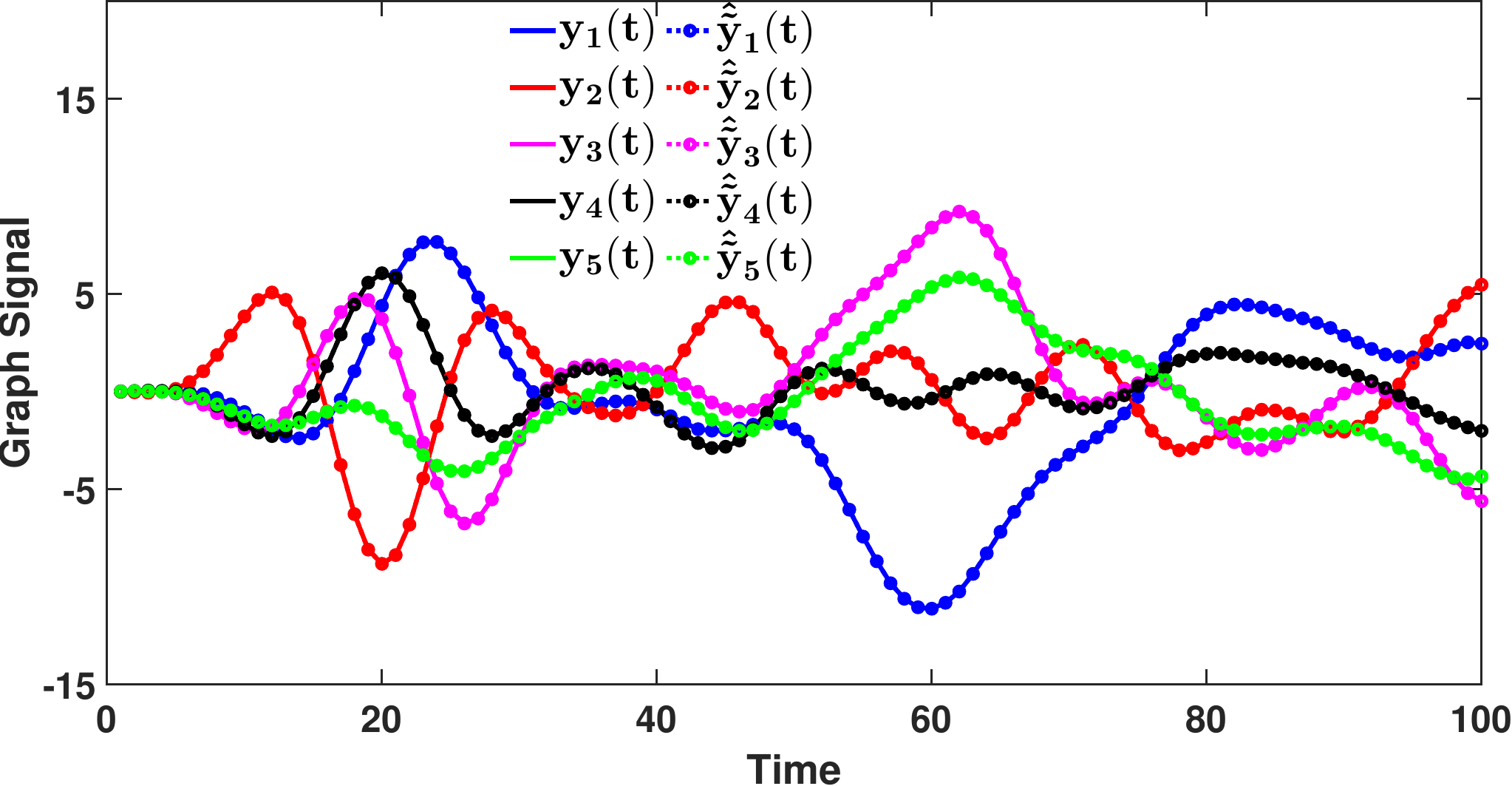} & $-192.69$ \\
        \cline{2-5}
        &&&& \\
        & $\tilde{\mathbf{B}}_{\text{samp}}$ &  \includegraphics[width=1.6in]{ICASSP_format/Images/Graph1_P_4_alpha_1.pdf} & \includegraphics[width=3.3in]{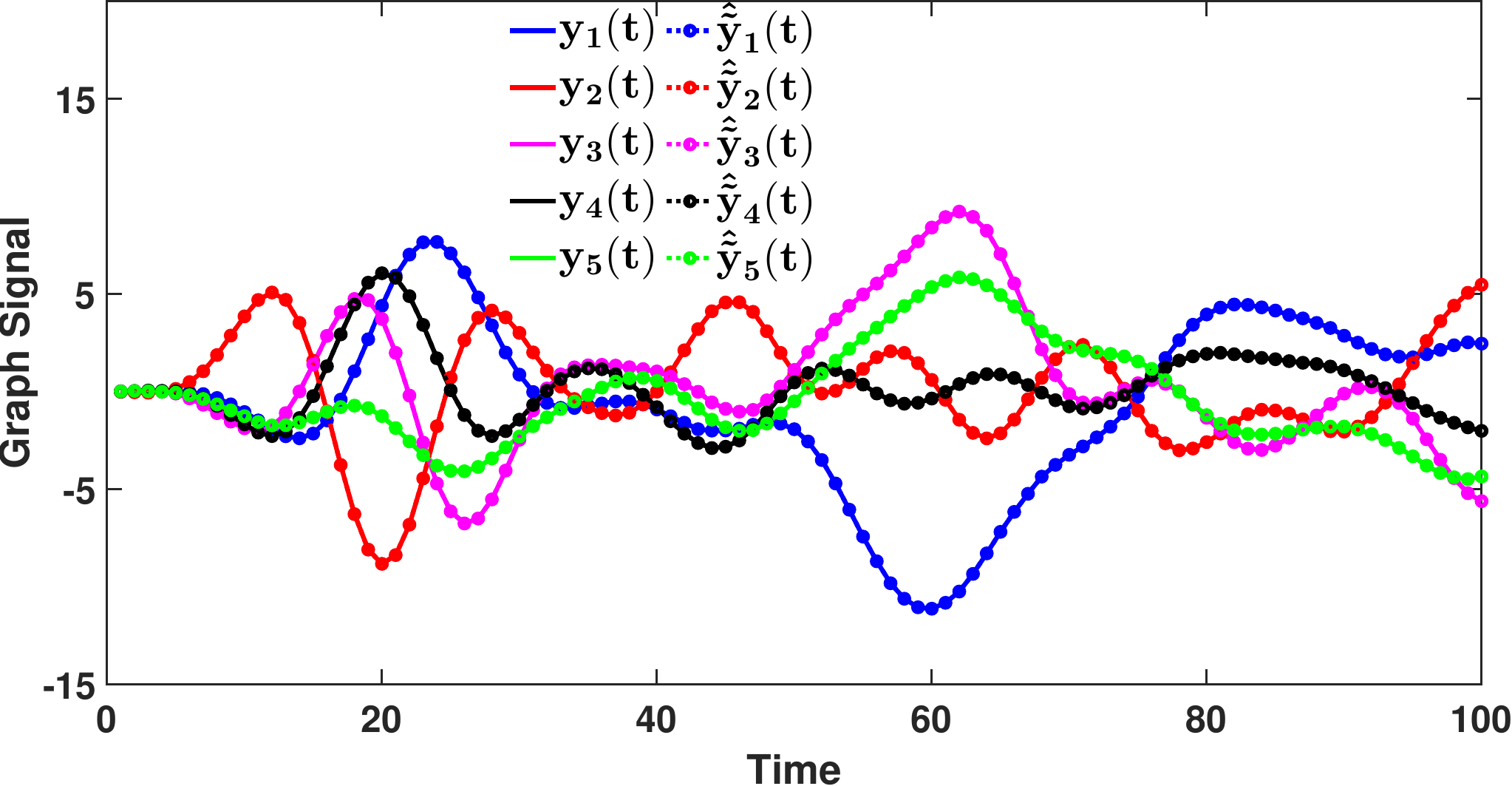} & $-222.04$ \\
        \hline

    \end{tabular}
    \egroup
\end{table*}

\begin{table*}
    \centering
    \caption{Results for an Erdos-R\`enyi graph with edge probability $p = 0.5$.}
    \label{tab:Res_Erdos_Renyi}
    \bgroup
    \renewcommand{\arraystretch}{1.5}
    \begin{tabular}{|m{0.2in}|m{0.25in}|m{1.6in}|m{3.3in}|m{0.75in}|}
        \hline
        $\boldsymbol{\varepsilon}$ & $\tilde{\mathbf{B}}$ & \textbf{Selected Nodes} & \textbf{Original and Reconstructed Signals} & \textbf{Normalized Error (in dB)} \\
        \hline
        \hline
        \multirow{4}{*}{0.03} &&&& \\
        &$\tilde{\mathbf{B}}_{\text{svd}}$ &  \includegraphics[width=1.6in]{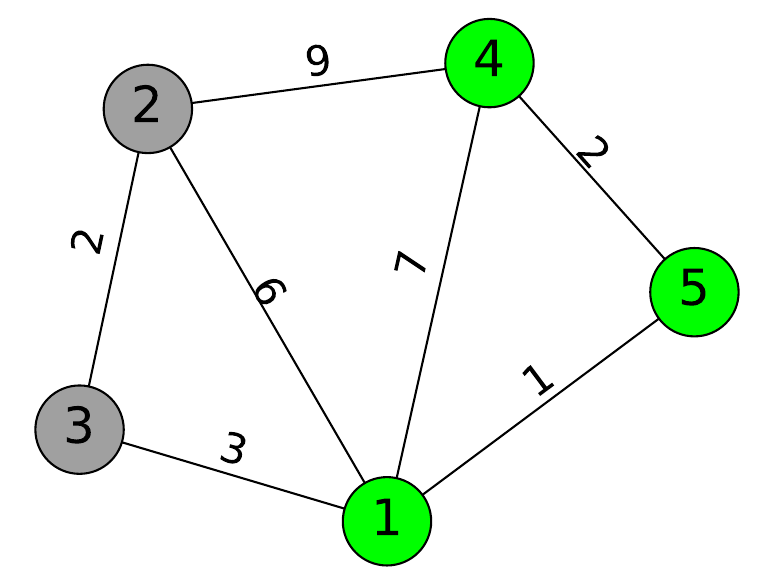} & \includegraphics[width=3.3in]{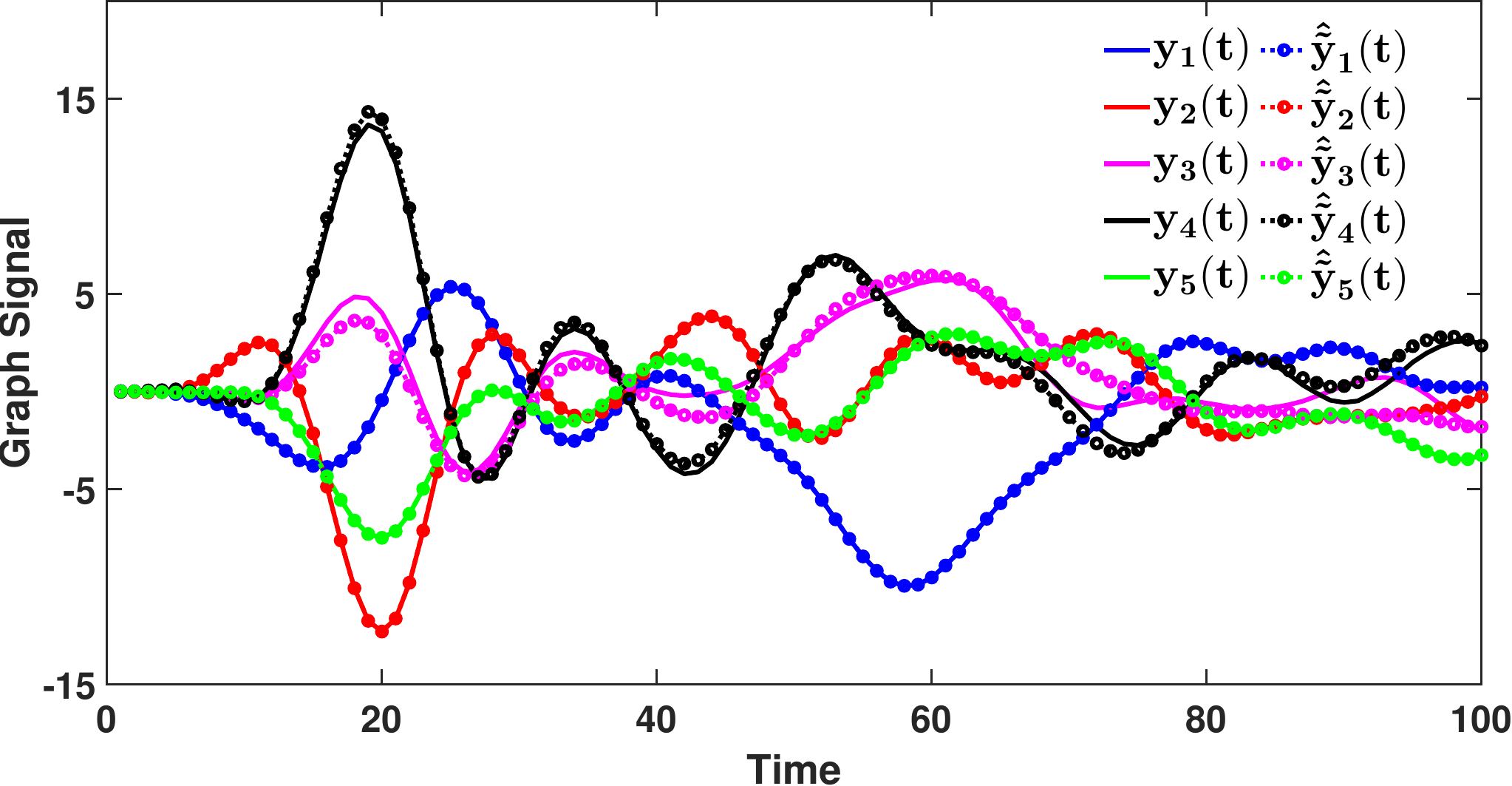} & $-17.81$ \\
        \cline{2-5}
        &&&& \\
        & $\tilde{\mathbf{B}}_{\text{samp}}$ &  \includegraphics[width=1.6in]{ICASSP_format/Images/Erdos_Renyi_Graph.pdf} & \includegraphics[width=3.3in]{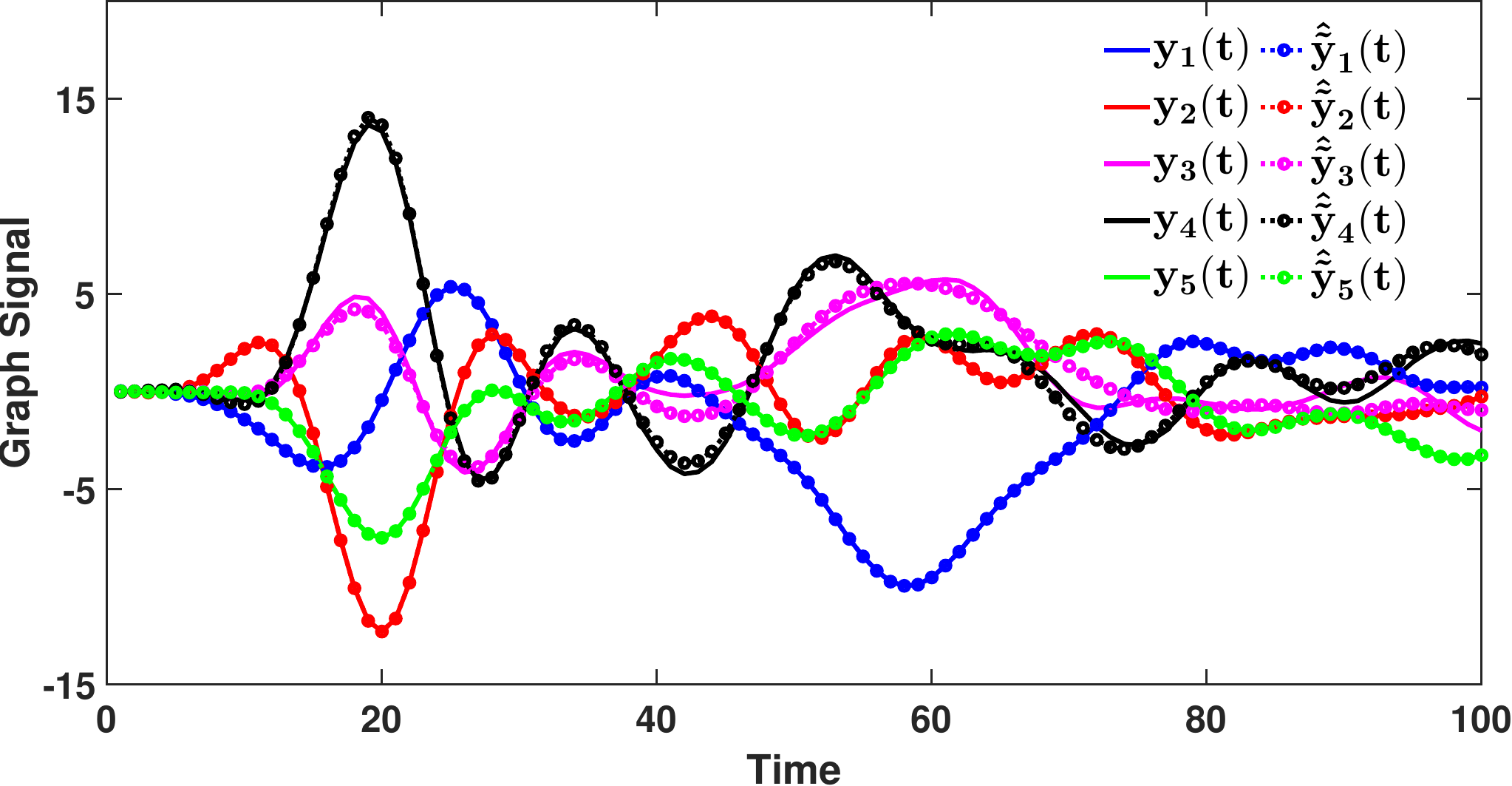} & $-19.89$ \\
        \hline

    \end{tabular}
    \egroup
\end{table*}

\begin{table*}
    \centering
    \caption{Results for a Complete graph with $5$ nodes.}
    \label{tab:Res_Comp_Graph}
    \bgroup
    \renewcommand{\arraystretch}{1.5}
    \begin{tabular}{|m{0.2in}|m{0.25in}|m{1.6in}|m{3.3in}|m{0.75in}|}
        \hline
        $\boldsymbol{\varepsilon}$ & $\tilde{\mathbf{B}}$ & \textbf{Selected Nodes} & \textbf{Original and Reconstructed Signals} & \textbf{Normalized Error (in dB)} \\
        \hline
        \hline
        \multirow{4}{*}{0.03} &&&& \\
        &$\tilde{\mathbf{B}}_{\text{svd}}$ &  \includegraphics[width=1.6in]{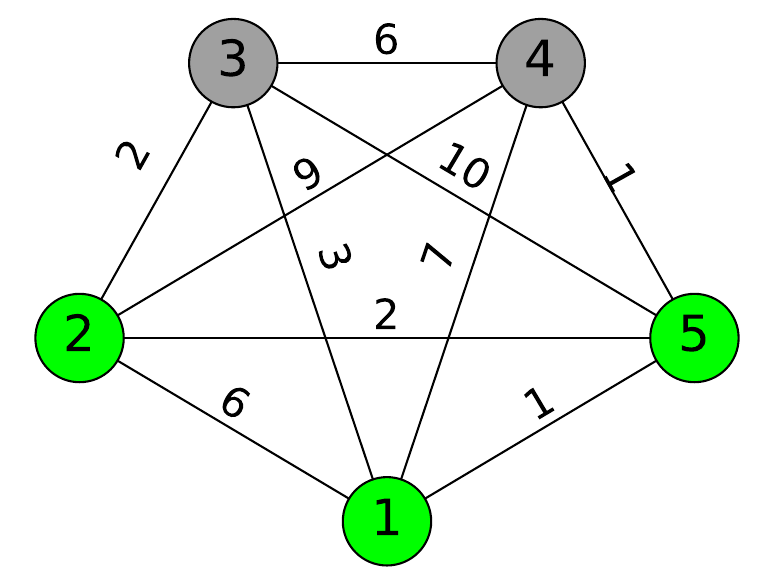} & \includegraphics[width=3.3in]{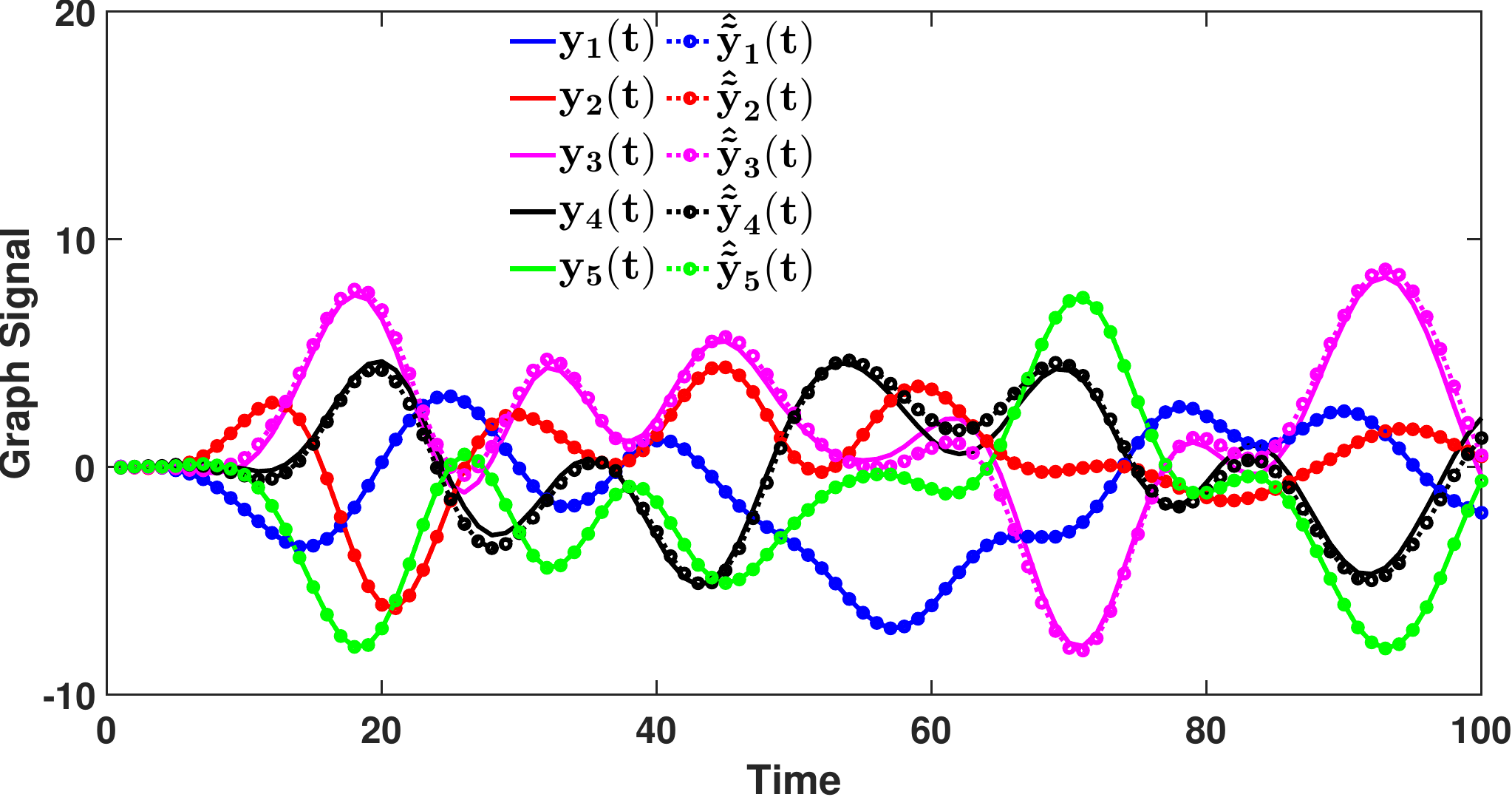} & $-20.15$ \\
        \cline{2-5}
        &&&& \\
        & $\tilde{\mathbf{B}}_{\text{samp}}$ &  \includegraphics[width=1.6in]{ICASSP_format/Images/Complete_Graph_P_3.pdf} & \includegraphics[width=3.3in]{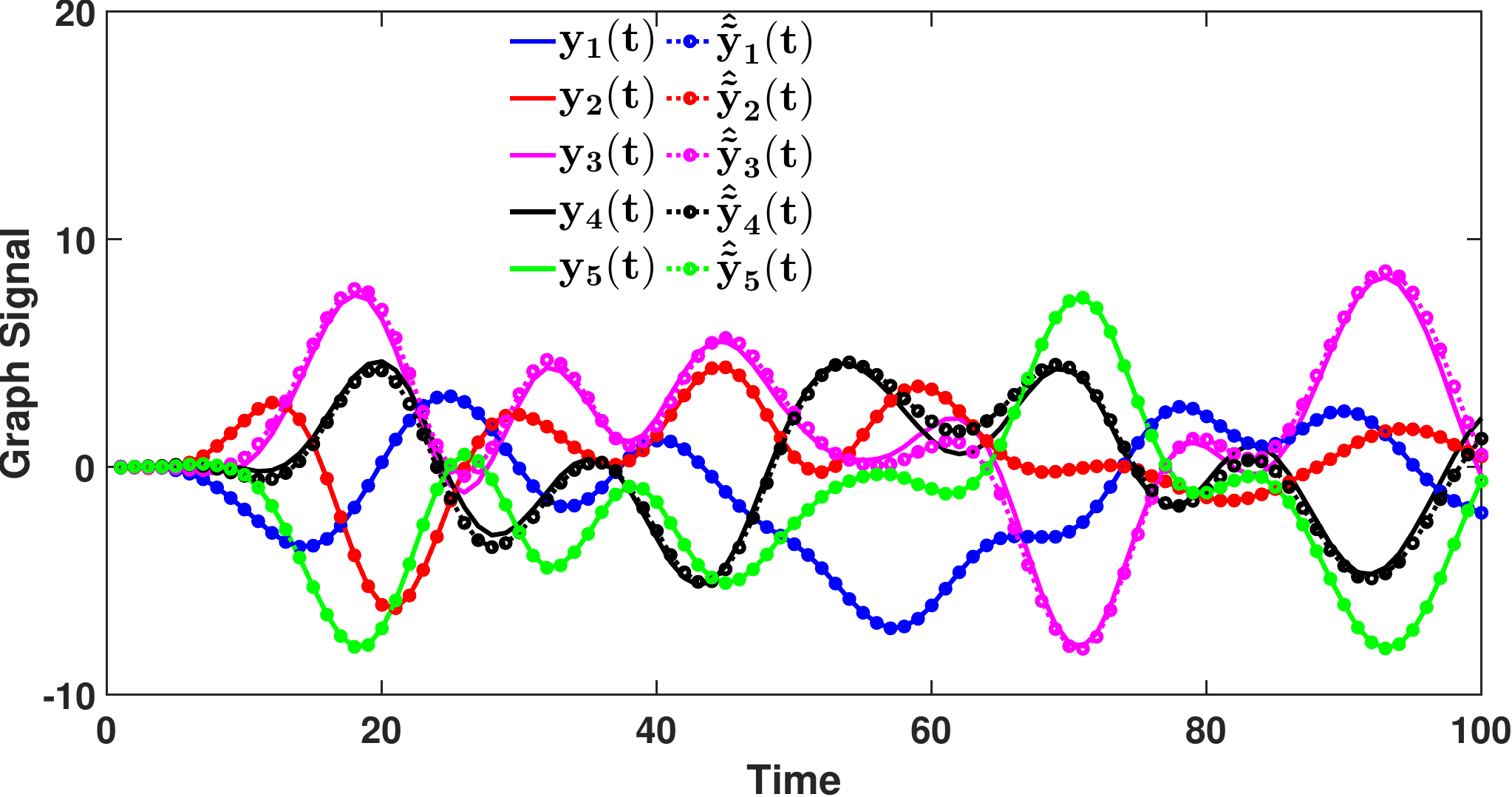} & $-20.52$ \\
        \hline

    \end{tabular}
    \egroup
\end{table*}

\begin{table*}
    \centering
    \caption{Results for a Bipartite graph with $5$ nodes.}
    \label{tab:Res_Bipartite}
    \bgroup
    \renewcommand{\arraystretch}{1.5}
    \begin{tabular}{|m{0.2in}|m{0.25in}|m{1.6in}|m{3.3in}|m{0.75in}|}
        \hline
        $\boldsymbol{\varepsilon}$ & $\tilde{\mathbf{B}}$ & \textbf{Selected Nodes} & \textbf{Original and Reconstructed Signals} & \textbf{Normalized Error (in dB)} \\
        \hline
        \hline
        \multirow{4}{*}{0.03} &&&& \\
        &$\tilde{\mathbf{B}}_{\text{svd}}$ &  \includegraphics[width=1.6in]{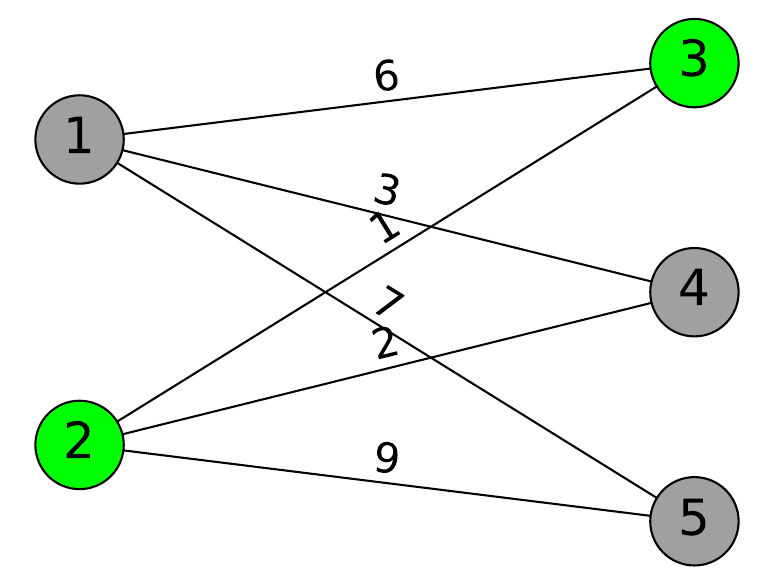} & \includegraphics[width=3.3in]{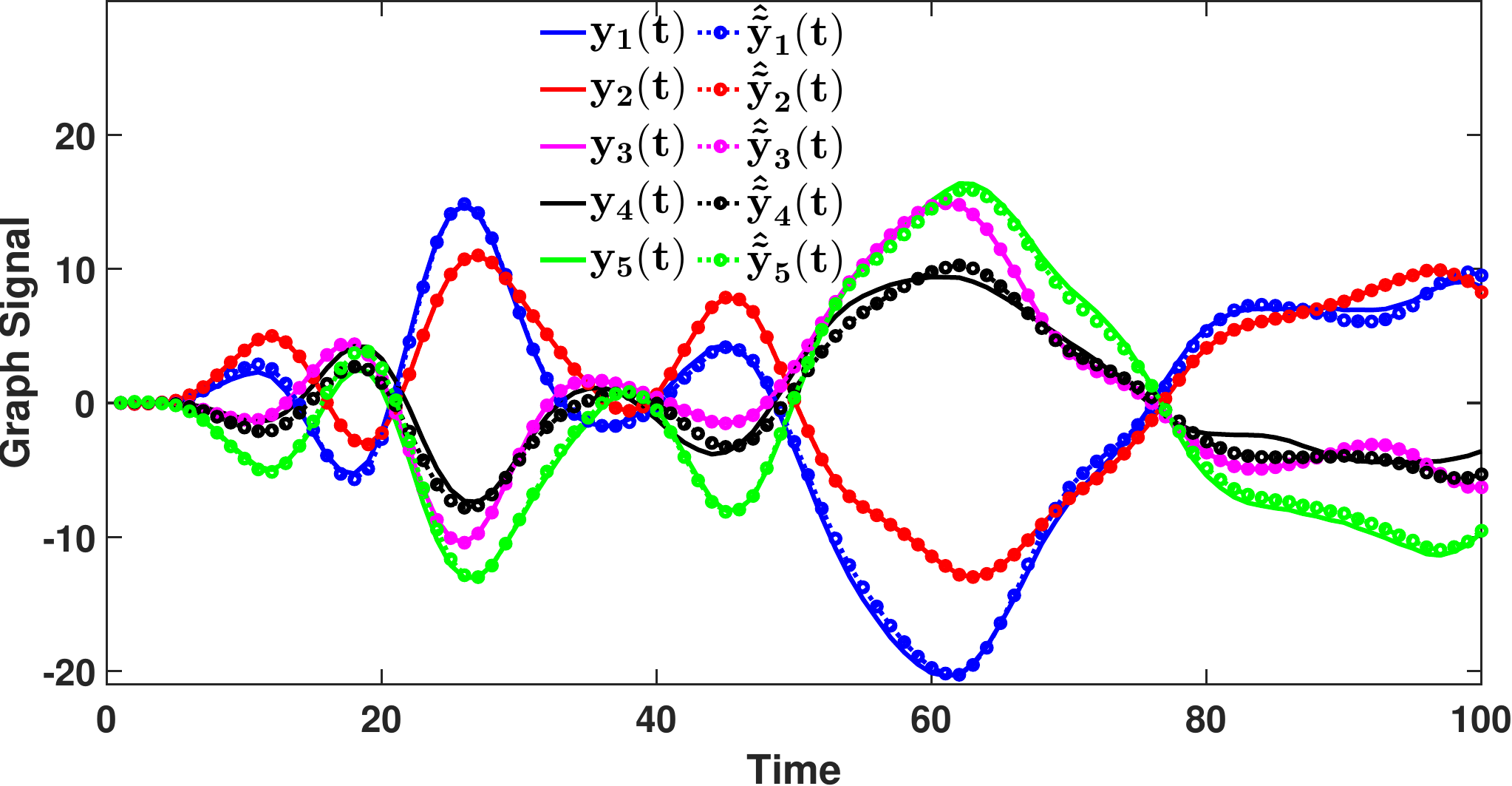} & $-22.44$ \\
        \cline{2-5}
        &&&& \\
        & $\tilde{\mathbf{B}}_{\text{samp}}$ &  \includegraphics[width=1.6in]{ICASSP_format/Images/Bipartite_Graph_P_2.pdf} & \includegraphics[width=3.3in]{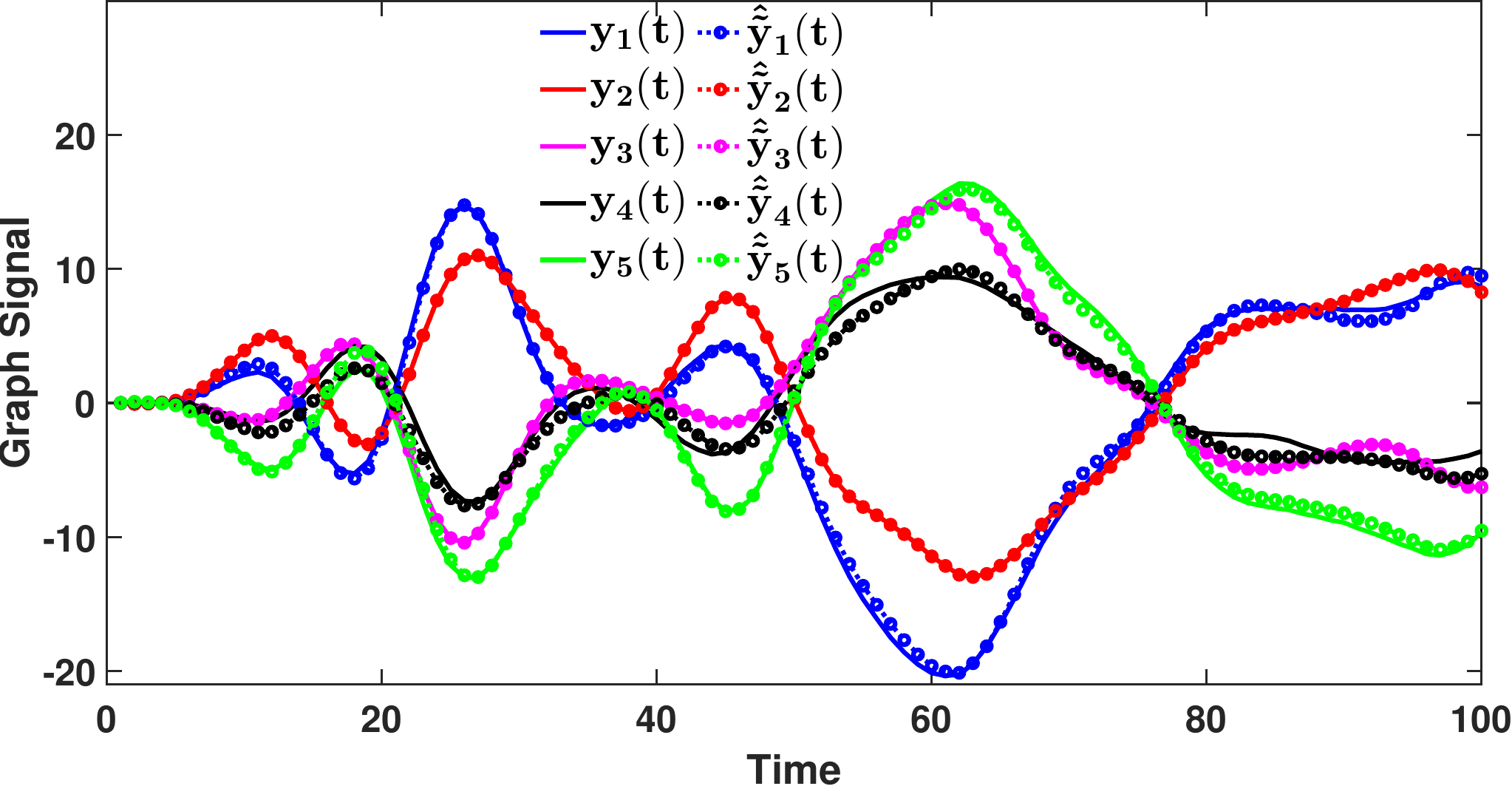} & $-22.23$ \\
        \hline

    \end{tabular}
    \egroup
\end{table*}

\bibliographystyle{IEEEtran}
\bibliography{refs,refs2,GSP}

\begin{thebibliography}{10}
\providecommand{\url}[1]{#1}
\csname url@samestyle\endcsname
\providecommand{\newblock}{\relax}
\providecommand{\bibinfo}[2]{#2}
\providecommand{\BIBentrySTDinterwordspacing}{\spaceskip=0pt\relax}
\providecommand{\BIBentryALTinterwordstretchfactor}{4}
\providecommand{\BIBentryALTinterwordspacing}{\spaceskip=\fontdimen2\font plus
\BIBentryALTinterwordstretchfactor\fontdimen3\font minus
  \fontdimen4\font\relax}
\providecommand{\BIBforeignlanguage}[2]{{%
\expandafter\ifx\csname l@#1\endcsname\relax
\typeout{** WARNING: IEEEtran.bst: No hyphenation pattern has been}%
\typeout{** loaded for the language `#1'. Using the pattern for}%
\typeout{** the default language instead.}%
\else
\language=\csname l@#1\endcsname
\fi
#2}}
\providecommand{\BIBdecl}{\relax}
\BIBdecl

\bibitem{eldar_2015sampling}
Y.~C. Eldar, \emph{Sampling Theory: Beyond Bandlimited Systems}.\hskip 1em plus
  0.5em minus 0.4em\relax Cambridge University Press, 2015.

\bibitem{mulleti_multiSubnyq_2021}
S.~Mulleti, K.~Lee, and Y.~C. Eldar, ``Sub-{N}yquist multichannel blind
  deconvolution,'' in \emph{Proc. IEEE Int. Conf. Acoust., Speech and Signal
  Process. (ICASSP)}, 2021, pp. 5454--5458.

\bibitem{ahmed_romberg_2015}
A.~{Ahmed} and J.~{Romberg}, ``Compressive multiplexing of correlated
  signals,'' \emph{IEEE Trans. Inf. Theory}, vol.~61, no.~1, pp. 479--498,
  2015.

\bibitem{Ahmed2020Compressive}
A.~Ahmed and J.~Romberg, ``Compressive sampling of ensembles of correlated
  signals,'' \emph{IEEE Trans. Inf. Theory}, vol.~66, no.~2, pp. 1078--1098,
  2020.

\bibitem{Joshi2009Sensor}
S.~Joshi and S.~Boyd, ``Sensor selection via convex optimization,'' \emph{IEEE
  Trans. Signal Process.}, vol.~57, no.~2, pp. 451--462, 2009.

\bibitem{MacKay1992Information}
D.~J.~C. MacKay, ``Information-based objective functions for active data
  selection,'' \emph{Neural Comput.}, vol.~4, no.~4, pp. 590--604, 07 1992.

\bibitem{Hashemi2021Randomized}
A.~Hashemi, M.~Ghasemi, H.~Vikalo, and U.~Topcu, ``Randomized greedy sensor
  selection: {L}everaging weak submodularity,'' \emph{IEEE Trans. Autom.
  Control}, vol.~66, no.~1, pp. 199--212, 2021.

\bibitem{Mulleti2020Fast}
S.~Mulleti, C.~Saha, H.~S. Dhillon, and Y.~C. Eldar, ``A fast-learning sparse
  antenna array,'' in \emph{IEEE Radar Conf.}, 2020, pp. 1--6.

\bibitem{Majumder2023Clustered}
K.~Majumder, S.~R.~B. Pillai, and S.~Mulleti, ``Clustered greedy algorithm for
  large-scale sensor selection,'' in \emph{IEEE Int. Conf. Acoust. Speech
  Signal Process. (ICASSP)}, 2023, pp. 1--5.

\bibitem{Leus2024Finding}
T.~Li and G.~Leus, ``Finding representative sampling subsets on graphs via
  submodularity,'' in \emph{IEEE Int. Conf. Acoust. Speech Signal Process.
  (ICASSP)}, 2024, pp. 9601--9605.

\bibitem{anis2014towards}
A.~Anis, A.~Gadde, and A.~Ortega, ``Towards a sampling theorem for signals on
  arbitrary graphs,'' in \emph{Int. Conf. Acoust. Speech Signal Process.
  (ICASSP)}.\hskip 1em plus 0.5em minus 0.4em\relax IEEE, 2014, pp. 3864--3868.

\bibitem{sakiyama2019eigendecomposition}
A.~Sakiyama, Y.~Tanaka, T.~Tanaka, and A.~Ortega, ``Eigendecomposition-free
  sampling set selection for graph signals,'' \emph{IEEE Trans. Signal
  Process.}, vol.~67, no.~10, pp. 2679--2692, 2019.

\bibitem{tanaka2020sampling}
Y.~Tanaka, Y.~C. Eldar, A.~Ortega, and G.~Cheung, ``Sampling signals on graphs:
  {F}rom theory to applications,'' \emph{IEEE Signal Process. Mag.}, vol.~37,
  no.~6, pp. 14--30, 2020.

\bibitem{yang2021efficient}
G.~Yang, L.~Yang, Z.~Yang, and C.~Huang, ``Efficient node selection strategy
  for sampling bandlimited signals on graphs,'' \emph{IEEE Tran. Signal
  Process.}, vol.~69, pp. 5815--5829, 2021.

\bibitem{Sawarkar2024Problems}
P.~Sawarkar, ``Problems in graph sampling and graph learning,'' Master's
  thesis, Dept. Comput. Sci. Eng., Indian Inst. Technol. Bombay, Mumbai, MH,
  India, 2024.

\bibitem{Xia2021Graph}
F.~Xia, K.~Sun, S.~Yu, A.~Aziz, L.~Wan, S.~Pan, and H.~Liu, ``Graph learning:
  {A} survey,'' \emph{IEEE Trans. Artif. Intell.}, vol.~2, no.~2, pp. 109--127,
  2021.

\bibitem{Egilmez2017Graph}
H.~E. Egilmez, E.~Pavez, and A.~Ortega, ``Graph learning from data under
  laplacian and structural constraints,'' \emph{IEEE J. Sel. Topics Signal
  Process.}, vol.~11, no.~6, pp. 825--841, 2017.

\bibitem{Kang2020Robust}
Z.~Kang, H.~Pan, S.~C.~H. Hoi, and Z.~Xu, ``Robust graph learning from noisy
  data,'' \emph{IEEE Trans. Cybern.}, vol.~50, no.~5, pp. 1833--1843, 2020.

\bibitem{chen2015signal}
S.~Chen, A.~Sandryhaila, J.~M. Moura, and J.~Kova{\v{c}}evi{\'c}, ``Signal
  recovery on graphs: Variation minimization,'' \emph{IEEE Trans. Signal
  Process.}, vol.~63, no.~17, pp. 4609--4624, 2015.

\bibitem{chen2016signal}
S.~Chen, R.~Varma, A.~Singh, and J.~Kova{\v{c}}evi{\'c}, ``Signal recovery on
  graphs: {F}undamental limits of sampling strategies,'' \emph{IEEE Trans.
  Signal Inf. Process. Net.}, vol.~2, no.~4, pp. 539--554, 2016.

\bibitem{qiu2017time}
K.~Qiu, X.~Mao, X.~Shen, X.~Wang, T.~Li, and Y.~Gu, ``Time-varying graph signal
  reconstruction,'' \emph{IEEE J. Sel. Topics Signal Process.}, vol.~11, no.~6,
  pp. 870--883, 2017.

\bibitem{ortega2018graph}
A.~Ortega, P.~Frossard, J.~Kova{\v{c}}evi{\'c}, J.~M. Moura, and
  P.~Vandergheynst, ``Graph signal processing: {O}verview, challenges, and
  applications,'' \emph{Proc. IEEE}, vol. 106, no.~5, pp. 808--828, 2018.

\bibitem{pesenson2008sampling}
I.~Pesenson, ``Sampling in {P}aley-{W}iener spaces on combinatorial graphs,''
  \emph{Trans. Amer. Math. Soc.}, vol. 360, no.~10, pp. 5603--5627, 2008.

\bibitem{dong2019learning}
X.~Dong, D.~Thanou, M.~Rabbat, and P.~Frossard, ``Learning graphs from data:
  {A} signal representation perspective,'' \emph{IEEE Signal Process. Mag.},
  vol.~36, no.~3, pp. 44--63, 2019.

\bibitem{mei2016signal}
J.~Mei and J.~M. Moura, ``Signal processing on graphs: {C}ausal modeling of
  unstructured data,'' \emph{IEEE Trans. Signal Process.}, vol.~65, no.~8, pp.
  2077--2092, 2016.

\bibitem{segarra2017network}
S.~Segarra, A.~G. Marques, G.~Mateos, and A.~Ribeiro, ``Network topology
  inference from spectral templates,'' \emph{IEEE Trans. Signal Inf. Process.
  Netw.}, vol.~3, no.~3, pp. 467--483, 2017.

\bibitem{thanou2017learning}
D.~Thanou, X.~Dong, D.~Kressner, and P.~Frossard, ``Learning heat diffusion
  graphs,'' \emph{IEEE Trans. Signal Inf. Process. Netw.}, vol.~3, no.~3, pp.
  484--499, 2017.

\bibitem{pasdeloup2017characterization}
B.~Pasdeloup, V.~Gripon, G.~Mercier, D.~Pastor, and M.~G. Rabbat,
  ``Characterization and inference of graph diffusion processes from
  observations of stationary signals,'' \emph{IEEE Trans. Signal Inf. Process.
  Netw.}, vol.~4, no.~3, pp. 481--496, 2017.

\bibitem{Eckart1936Approximation}
C.~Eckart and G.~Young, ``The approximation of one matrix by another of lower
  rank,'' \emph{Psychometrika}, vol.~1, no.~3, pp. 211--218, 1936.

\end{thebibliography}

\end{document}